\documentclass[twocolumn]{autart}    

\usepackage{amsmath,amsfonts,amssymb}
\usepackage{mathrsfs}
\usepackage{mathtools}
\usepackage{cuted}  
\usepackage{array}
\usepackage{float}
\usepackage{extarrows}
\usepackage{cite}
\usepackage{url}
\usepackage{color}
\usepackage{comment}
\usepackage[table]{xcolor}
\usepackage[colorlinks,linkcolor=orange,citecolor=blue]{hyperref}
\usepackage{algorithmic}
\usepackage{graphicx}
\usepackage{textcomp}
\usepackage[font=footnotesize]{caption}
\usepackage[font=footnotesize]{subcaption}
\usepackage{stfloats}
\usepackage{epstopdf}
\usepackage{arydshln}
\usepackage{ifthen}
\usepackage{pict2e}
\usepackage[T1]{fontenc}
\usepackage{enumitem}
\usepackage{tikz}

\allowdisplaybreaks[1] 

\DeclareGraphicsExtensions{.pdf,.png}
\DeclareMathOperator{\blkdiag}{blkdiag}
\DeclareMathOperator{\diag}{diag}

\DeclareMathOperator{\trace}{trace}

\DeclareMathOperator{\vect}{vec}
\DeclareMathOperator{\range}{range}
\DeclareMathOperator{\cone}{cone}
\DeclareMathOperator{\spec}{spec}

\newtheorem{theorem}{Theorem}

\newcommand{\dd}{\mathrm{d}} 
\renewcommand{\t}{^{\mbox{\tiny\sf T}}} 

\newcommand{\panos}[1]{{\color{red}[Panos: #1]}}
\newcommand{\feng}[1]{{\color{teal}[Fengjiao: #1]}}

\newcommand{\R}{\mathbb{R}}

\def\send#1#2{\stackrel{#1}{\hbox to #2{\rightarrowfill}}}
\def\-{\!\!\!\!\!-}

 \def\qed{ \rule{.1in}{.1in}}

\newcommand{\rank}{{\rm rank\;}}

\newtheorem{lemma}{Lemma}
\newtheorem{remark}{Remark}
\newtheorem{proposition}{Proposition}
\newtheorem{corollary}{Corollary}

\newtheorem{assumption}{Assumption}

\newtheorem{problem}{Problem}

\def\qed{ \rule{.1in}{.1in}}

\def\R{{\rm I\!R}} 

\newcounter{seqn}[equation]
\def\theseqn{\arabic{equation}\alph{seqn}}

\def\endseqn{\eqno \@seqnnum
$$\ignorespaces}
\def\@seqnnum{(\theseqn)}

\newskip\mcentering \mcentering=0pt plus 1000pt minus 1000pt

\def\meqalignno#1{
\halign to\displaywidth{
    \hbox to 0pt{\kern\displaywidth\llap{$##$}\hss}\tabskip=\mcentering
    &\hfil$\displaystyle{##}$\tabskip=\mcentering
   &&$\displaystyle{{}##}$\hfil\tabskip=\mcentering
    \crcr
    #1\crcr}}

\def\dspace{\multiply\normalbaselineskip 150
		  \divide\normalbaselineskip 100 \normalbaselines
		  \csname @@normalbaselineskip\endcsname\normalbaselineskip}
\def\sspace{\multiply\normalbaselineskip 200
		 \divide\normalbaselineskip 300 \normalbaselines
		 \csname @@normalbaselineskip\endcsname\normalbaselineskip}
\def\sdspace{\multiply\normalbaselineskip 160
		 \divide\normalbaselineskip 150 \normalbaselines
		 \csname @@normalbaselineskip\endcsname\normalbaselineskip}

\def\@{\tilde}

\def\3dot#1{\buildrel\textstyle...\over#1}

\renewcommand{\R}{\mathbb{R}}

\makeatletter
\newcommand{\ostar}{\mathbin{\mathpalette\make@circled\star}}
\newcommand{\make@circled}[2]{%
\ooalign{$\m@th#1\smallbigcirc{#1}$\cr\hidewidth$\m@th#1#2$\hidewidth\cr}%
}
\newcommand{\smallbigcirc}[1]{%
\vcenter{\hbox{\scalebox{0.77778}{$\m@th#1\bigcirc$}}}%
}
\makeatother

\begin{document}

\begin{frontmatter}

\title{Reachability and Controllability Analysis of the State Covariance 
for Linear Stochastic Systems\thanksref{footnoteinfo}} 

\thanks[footnoteinfo]{Corresponding author F. Liu. 
}

\author[FSU]{Fengjiao Liu}\ead{fliu@eng.famu.fsu.edu},    
\and
\author[GT]{Panagiotis Tsiotras}\ead{tsiotras@gatech.edu}               

\address[FSU]{Department of Electrical and Computer Engineering,
FAMU-FSU College of Engineering, Tallahassee, FL 32310, USA}   
\address[GT]{School of Aerospace Engineering, Georgia Institute of Technology, Atlanta, GA 30332, USA}  


\begin{abstract}                          
This paper studies the set of terminal state covariances that are reachable over a finite time horizon from a given initial state covariance for a linear stochastic system subject to additive noise. 
For discrete-time systems, a complete characterization of the set of reachable state covariances is given. 
For continuous-time systems, we present an upper bound on the set of reachable state covariances. 
Moreover, for both linear discrete-time and continuous-time systems, necessary and sufficient conditions are provided for the controllability of the state covariance over a finite horizon. 
\end{abstract}

\end{frontmatter}


\section{Introduction}

For the past forty years, covariance control theory has emerged as a tool for addressing the growing need to quantify and control the uncertainty in dynamical systems. 
The study of covariance control theory first addressed the \emph{infinite-horizon} state covariance assignment problem~\cite{collins1987theory, hotz1987covariance}, which aims to find the set of all admissible state-state covariances of a linear stochastic system subject to additive noise. 
In the same works, the parameterization of all state-feedback controllers for attaining an assignable state covariance is also presented. 
The optimal control law for achieving an admissible steady-state covariance with minimum control energy is provided in a closed form in~\cite{grigoriadis1997minimum}.

In recent years, covariance steering theory, which aims to steer the state covariance from a given initial covariance to a desired terminal covariance over a \emph{finite horizon}, has gained popularity in a variety of control and motion planning problems~\cite{ridderhof2018uncertainty, okamoto2019optimal, yin2022trajectory, knaup2023safe, zheng:tro2024}. 
With these applications in mind, several optimal covariance steering problems have been studied for both continuous-time and discrete-time stochastic systems. 
For a continuous-time linear stochastic system with a quadratic cost, when the noise coefficient matrix coincides with the control coefficient matrix, the optimal control law can be derived in closed form~\cite{chen2016I, chen2018III, ciccone2020regularized}. 
When the noise coefficient matrix differs from the control coefficient matrix, the optimal control law can be computed by recasting the optimal covariance steering problem as a semi-definite program~\cite{liu2024add, liu2024mult, chen2016II}. 
In addition to the first two moments, the question of how to optimally steer the  whole state density of a stochastic system between two prescribed endpoint densities with minimum control energy has also been studied~\cite{chen2017optimal, caluya2022wasserstein}. 
%
In the basic discrete-time version, the optimal control policy has been derived for steering the state covariance of a discrete-time linear stochastic system with different noise models and objective functions~\cite{balci2022exact, liu2024disc, balci2023covariance} and different types of constraints (e.g., chance constraints)~\cite{bakolas2018finite, okamoto2018optimal, renganathan2023distributionally, rapakoulias2023discrete}. 
Furthermore, the system matrices may be subject to unknown parameters~\cite{knaup2023covariance} or can be completely unknown~\cite{pilipovsky2023data}. 
General distribution steering problems in discrete time have been tackled using either characteristic functions or power moments~\cite{sivaramakrishnan2021distribution, wu2023Liouville, wu2023FokkerPlanck}. 
Multi-agent covariance and distribution steering problems are explored in~\cite{saravanos2021distributed, terpin2024dynamic}.

Despite the extensive research on optimal covariance steering theory, the more fundamental question regarding which terminal state covariances can be reached from a given initial state covariance via a proper control law remains largely unanswered, except for some special cases. 
One special case, where a complete analysis is available, is the controllability of the continuous-time state covariance.
This case addresses the problem under which (sufficient) conditions a positive-definite terminal state covariance can be reached from a given initial positive-definite state covariance. 
Specifically, as it was shown in~\cite{brockett2012notes}, for a continuous-time linear time-invariant (LTI) system $\dot{x}(t) = A x(t) + B u(t)$, its state covariance is controllable on a finite time interval if the matrix pair $(A, B)$ is controllable. 
This result still holds even when the LTI system is subject to additive noise~\cite{mahmudov2000controllability}. 
Similarly, for a linear time-varying stochastic system $\dd x(t) = \big(A(t) x(t) + B(t) u(t)\big) \dd t + D(t) \, \dd w(t)$, its state covariance is controllable on the time interval $[0, T]$ if, for all $s \in [0, T)$, the matrix pair $\big(A(t), B(t)\big)$ is controllable on $[s, T]$~\cite{mahmudov2001controllability}. 
Another special case where controllability has been analyzed
is for the discrete-time one-dimensional linear stochastic system, where it has been shown~\cite{wu2023Liouville} that the
state covariance (and higher power moments) is controllable in a finite time, provided that the control input is independent of the state. 
%

In this paper, we study the more general problem of the reachability of the terminal state covariance for linear stochastic systems, and we give necessary and sufficient conditions for the controllability of the state covariance over a finite horizon.

\emph{Contributions:} 
First, we provide a complete characterization of the set of reachable state covariances for a \emph{discrete-time} linear stochastic system with and without additive noise (Theorem~\ref{thm:reach-no-noise}, Theorem~\ref{thm:reach-noise}, and Theorem~\ref{thm:reach-noise-2}).
Second, we present an upper bound on the reachable set of terminal state covariances for a \emph{continuous-time} linear stochastic system corrupted by additive noise (Theorem~\ref{thm:reach-noise-bd}). 
Third, we establish necessary and sufficient conditions for the finite-horizon controllability of the state covariance for \emph{both discrete-time} and \emph{continuous-time} linear stochastic systems in the presence of additive noise (Theorem~\ref{thm:contr-noise} and Theorem~\ref{thm:contr-noise-cont}). 
In the process of our analysis, we also derive necessary and sufficient conditions for a Riccati differential equation to admit a unique solution on a given time interval (Theorem~\ref{thm:pi-exist}), which may be of independent interest. 
%
%
To the best of our knowledge, this is the first work addressing the reachability of the state covariance over a finite horizon, which complements the assignability results (over an infinite horizon) by Skelton et al.~\cite{collins1987theory, hotz1987covariance}.

The rest of the paper is organized as follows. 
The problems of interest are formulated in Section~\ref{sec:problem}. 
The discrete-time  and continuous-time results are presented in Section~\ref{sec:discrete} and Section~\ref{sec:continuous}, respectively. 
A numerical example demonstrating the theory is given in Section~\ref{sec:example}. 
To help the exposition, some auxiliary lemmas used in the proofs of the main results in the paper are postponed to the Appendix. 


\section{Problem Formulation} \label{sec:problem}

Consider the discrete-time linear stochastic system
\begin{equation} \label{sys:disc-time}
x_{k+1} = A_{k} x_{k} + B_{k} u_{k} + D_{k} w_{k}, \quad k = 0, 1, 2, \dots
\end{equation}
where $x_{k} \in \R^{n}$ is the state, $u_{k} \in \R^{p}$ is the control input, 
and $\{w_{k}\}_{k\ge 0}$ is a sequence of independent, square-integrable $q$-dimensional random noise vectors with $\mathbb{E}\left[w_{k}\right] = 0$ and $\mathbb{E}\left[w_{k} w_{k}\t\right] = I_{q}$, 
and $A_{k} \in \R^{n \times n}$, $B_{k} \in \R^{n \times p}$, and $D_{k} \in \R^{n \times q}$ are known  coefficient matrices.

Assume that the initial state $x_{0}$ has a finite positive semi-definite covariance matrix $\Sigma_{0} \succeq 0$. 
At any given time $k = 1, 2, \dots$, a state covariance $\bar{\Sigma}_{k}$ is said to be \emph{reachable from} $\Sigma_{0}$ if there exists a finite-expected-energy control sequence $\{u_0,u_1,\ldots,u_{k-1}\}$ such that, at each time step $i = 0, 1, \dots, k-1$, $u_i$ depends only on $i$ and on the past history of the states $x_{0}, x_{1}, \dots, x_{i}$, and drives the state covariance of \eqref{sys:disc-time} from $\Sigma_{0}$ at time $0$ to $\bar{\Sigma}_{k}$ at time $k$. 
In particular, the state covariance of \eqref{sys:disc-time} is \emph{controllable from time $0$ to $k$} if, for any pair of matrices $\bar{\Sigma}_{0}, \bar{\Sigma}_{k} \succeq 0$, there exists such a control sequence 
$\{u_0,u_1,\ldots,u_{k-1}\}$
that steers the state covariance of the system from $\Sigma_{0} = \bar{\Sigma}_{0}$ to $\Sigma_{k} = \bar{\Sigma}_{k}$.

\begin{problem} \label{prb:disc-reach}
Find the set of reachable state covariances of the system~\eqref{sys:disc-time} at time $k = 1, 2, \dots$ starting from a given initial state covariance $\Sigma_{0} \succeq 0$ at time $0$. 
\end{problem}

\begin{problem} \label{prb:disc-ctrl}
Find necessary and sufficient conditions for the state covariance of the system~\eqref{sys:disc-time} to be controllable from time $0$ to $k = 1, 2, \dots$. 
\end{problem}


We will also consider the continuous-time linear stochastic system
\begin{equation} \label{sys:cont-time}
\dd x(t) = A(t) x(t) \, \dd t + B(t) u(t) \, \dd t + D(t) \, \dd w(t),
\end{equation}
for $0 \leq t \leq T$, where $x(t) \in \R^{n}$ is the state, $u(t) \in \R^{p}$ is the control input, 
$w(t) \in \R^{q}$ is an independent, square-integrable, martingale noise signal with $\dd \mathbb{E} \big[w(t) w(t)\t \big] = I_{q} \, \dd t$, and
$A(t) \in \R^{n \times n}$, $B(t) \in \R^{n \times p}$, and $D(t) \in \R^{n \times q}$ are known coefficient matrices.

Similarly to the discrete-time case, we assume that the initial state $x(0)$ has a finite positive semi-definite covariance matrix $\Sigma_{0} \succeq 0$. 
At any given time $T > 0$, a state covariance $\bar{\Sigma}_{T}$ is said to be \emph{reachable} from $\Sigma_{0}$ if there exists a finite-expected-energy control 
$u$ that, for each time $t \in [0, T]$, depends only on $t$ and on the past history of the states $\{x(s) : 0 \leq s \leq t\}$, and drives the state covariance of \eqref{sys:cont-time} from $\Sigma_{0}$ at time $0$ to $\bar{\Sigma}_{T}$ at time $T$. 
In particular, the state covariance of \eqref{sys:cont-time} is \emph{controllable} from time $0$ to $T$ if, for any pair of matrices $\bar{\Sigma}_{0}, \bar{\Sigma}_{T} \succ 0$, there exists such a control $u$ that steers the state covariance of the system from $\Sigma(0) = \bar{\Sigma}_{0}$ to $\Sigma(T) = \bar{\Sigma}_{T}$.

\begin{problem} \label{prb:cont-reach}
Find the set of reachable state covariances of the system~\eqref{sys:cont-time} at time $T > 0$ starting from a given initial state covariance $\Sigma_{0} \succeq 0$ at time $0$. 
\end{problem}

\begin{problem} \label{prb:cont-ctrl}
Find necessary and sufficient conditions for the state covariance of the system~\eqref{sys:cont-time} to be controllable from time $0$ to $T > 0$. 
\end{problem}

In this paper, we will provide complete answers to Problem~\ref{prb:disc-reach}, Problem~\ref{prb:disc-ctrl}, and Problem~\ref{prb:cont-ctrl}, as well as a partial answer to Problem~\ref{prb:cont-reach}.

For clarity, some notation used in this paper is briefly summarized below. 
Let $M$ be an $n \times m$ real-valued matrix. 
The column span of $M$ is denoted by $\range M$, which is a (linear) subspace in $\R^{n}$. 
Let $\mathcal{Z}$ be a subspace in $\R^{n}$. 
The \emph{orthogonal complement} of $\mathcal{Z}$, 
denoted by $\mathcal{Z}^{\perp}$,
is the set of all vectors in $\R^{n}$ that are orthogonal to every vector in $\mathcal{Z}$. 
It can be verified that $\mathcal{Z}^{\perp}$ is also a subspace in $\R^{n}$. 
The \emph{orthogonal projection} onto the subspace $\mathcal{Z}$ is the projection along $\mathcal{Z}^{\perp}$ onto $\mathcal{Z}$. 
Let $Z$ be any matrix whose columns form a basis of $\mathcal{Z}$. 
Then, the orthogonal projection matrix onto $\mathcal{Z}$ is given by $Z (Z\t Z)^{-1} Z\t$. 
For an $n \times n$ symmetric matrix $M$, its \emph{square root} $M^{\frac{1}{2}}$ is given by 
$M^{\frac{1}{2}} = P \Lambda^{\frac{1}{2}} P\t$, where $\Lambda^{\frac{1}{2}} = \diag \left[\sqrt{\lambda_{1}}, \dots, \sqrt{\lambda_{j}}, i \sqrt{\lambda_{j+1}}, \dots, i \sqrt{\lambda_{k}}, 0, \dots, 0\right]$ and $i$ is the imaginary unit, 
where $P$ is an orthogonal matrix and 
$\Lambda = \diag \left[\lambda_{1}, \dots, \lambda_{j}, -\lambda_{j+1}, \dots, -\lambda_{k}, 0, \dots, 0\right]$ 
is a real diagonal matrix with $\lambda_{1}, \dots, \lambda_{k} > 0$ 
such that $M = P \Lambda P\t$.


\vspace{-1mm}
\section{Discrete-Time Case} \label{sec:discrete}

In this section, we first show that it suffices to consider the class of randomized state-feedback controls for the system~\eqref{sys:disc-time}. 
Then, we characterize the set of all reachable state covariances for the special case when \eqref{sys:disc-time} has no noise, that is, $D_{k} \equiv 0$. 
Next, we extend the result to the general case when $D_{k} \neq 0$. 
Lastly, from the reachability of the state covariance, we derive a necessary and sufficient condition for the controllability of the state covariance of~\eqref{sys:disc-time} over a finite horizon.

\subsection{Randomized State-Feedback Controls}

First, we introduce a result from \cite{zhang2005schur} for characterizing a positive semi-definite matrix, which will be used multiple times in this paper.

\begin{proposition} \label{prp:pos-semi-defn}
Let $M$ be a symmetric matrix partitioned as 
\begin{equation*}
M = 
\begin{bmatrix}
M_{1} & M_{2} \\
M_{2}\t & M_{3}
\end{bmatrix}, 
\end{equation*}
where $M_{1}$ and $M_{3}$ are square. 
Then, $M \succeq 0$ if and only if $M_{1} \succeq 0$, $\range M_{2} \subseteq \range M_{1}$, and $M_{3} - M_{2}\t M_{1}^{-} M_{2} \succeq 0$, where $M_{1}^{-}$ is a generalized inverse (not necessarily unique) of $M_{1}$ satisfying $M_{1} M_{1}^{-} M_{1} = M_{1}$. 
\end{proposition}

Let the \emph{generalized Schur complement} of the block $M_{1}$ in the matrix $M$ be denoted by $M/M_{1} \triangleq M_{3} - M_{2}\t M_{1}^{-} M_{2}$. 
Clearly, $M_{1}^{-}$ can be chosen as the (unique) 
Moore–Penrose inverse $M_{1}^{+}$ of $M_{1}$. 
We will use Proposition~\ref{prp:pos-semi-defn} to show that we only need to consider randomized state feedback controls to steer the state covariance of~\eqref{sys:disc-time}.

Let $\Sigma_{k}$ denote the covariance of the state $x_{k}$ at time $k$ in \eqref{sys:disc-time}. 
The state covariance then satisfies
\begin{multline} \label{eqn:stt-covar}
\Sigma_{k+1} = 
A_{k} \Sigma_{k} A_{k}\t + B_{k} \Sigma^{ux}_{k} A_{k}\t + A_{k} \Sigma^{ux \mbox{\tiny\sf T}}_{k} B_{k}\t 
\\*
+ B_{k} \Sigma^{uu}_{k} B_{k}\t + D_{k} D_{k}\t, 
\end{multline}
where $\Sigma^{ux}_{k}$ is the cross covariance of $u_{k}$ and $x_{k}$, and $\Sigma^{uu}_{k}$ is the covariance of $u_{k}$. 
They are both treated as ``control'' terms in equation~\eqref{eqn:stt-covar}. 
Since the covariance matrix of the stacked vector $[x_{k}\t ~~ u_{k}\t]\t$ is 
$\begin{bmatrix}
\Sigma_{k} & \Sigma^{ux \mbox{\tiny\sf T}}_{k} \\
\Sigma^{ux}_{k} & \Sigma^{uu}_{k}
\end{bmatrix}$, 
which is always positive semi-definite, it follows from Proposition~\ref{prp:pos-semi-defn} that there exists a deterministic matrix $F_{k} \in \R^{p \times n}$, which may depend on the distributions of $u_{k}$ and $x_{k}$, such that $\Sigma^{ux}_{k} = F_{k} \Sigma_{k}$ and $\Sigma^{uu}_{k} \succeq \Sigma^{ux}_{k} \Sigma_{k}^{+} \Sigma^{ux \mbox{\tiny\sf T}}_{k} = F_{k} \Sigma_{k} F_{k}\t$. 
Thus, \eqref{eqn:stt-covar} can be written equivalently as 
\begin{equation} \label{eqn:stt-covar-fb}
\Sigma_{k+1} 
\hspace{-0.3mm} = \hspace{-0.3mm}
\big(A_{k} + B_{k} F_{k}\big) \Sigma_{k} \big(A_{k} + B_{k} F_{k}\big)\t \hspace{-0.1mm} + B_{k} V_{k} B_{k}\t + D_{k} D_{k}\t, 
\end{equation}
where $V_{k} = \Sigma^{uu}_{k} - F_{k} \Sigma_{k} F_{k}\t \succeq 0$. 
Hence, we have reached the following conclusion.

\begin{proposition} \label{prp:rand-stt-fb}
At time $K = 1, 2, \dots$, if a terminal state covariance $\bar{\Sigma}_{K}$ of the discrete-time system~\eqref{sys:disc-time} is reachable from a given initial state covariance $\Sigma_{0}$, then, $\bar{\Sigma}_{K}$ is also reachable via a randomized state feedback control $u_{k} = F_{k} x_{k} + \nu_{k}$, 
for some $F_{k} \in \R^{p \times n}$, 
where $\nu_{k} \in \R^{p}$, ($k = 0, 1, \dots, K-1$) is a sequence of independent, square-integrable random vectors with $\mathbb{E}\left[\nu_{k}\right] = 0$ and $\mathbb{E}\left[\nu_{k} \nu_{k}\t\right] \succeq 0$. 
\end{proposition}

\noindent {\bf Proof:} 
Apply the control $u_{k} = F_{k} x_{k} + \nu_{k}$ to system~\eqref{sys:disc-time}, where $\mathbb{E}\left[\nu_{k} \nu_{k}\t\right] = V_{k} \succeq 0$. 
We then obtain that $x_{k+1} = (A_{k} + B_{k} F_{k}) x_{k} + B_{k} \nu_{k} + D_{k} w_{k}$. 
Since $\mathbb{E}\left[\nu_{k}\right] = 0$, 
$\mathbb{E}\left[w_{k}\right] = 0$, 
$\mathbb{E}\left[\nu_{k} \nu_{k}\t\right] = V_{k}$, and 
$\mathbb{E}\left[w_{k} w_{k}\t\right] = I_{q}$, 
it follows that the state covariance of the closed-loop system satisfies \eqref{eqn:stt-covar-fb}. 
In light of the equivalence of the state covariance equations \eqref{eqn:stt-covar} and \eqref{eqn:stt-covar-fb}, Proposition~\ref{prp:rand-stt-fb} holds. 
\hfill $\qed$


\subsection{Reachability of the State Covariance Without Noise}

Without additive noise in \eqref{sys:disc-time}, we have $D_{k} \equiv 0$. 
In light of \eqref{eqn:stt-covar-fb}, the state covariance of \eqref{sys:disc-time} in this case propagates according to
\begin{equation} \label{eqn:stt-covar-no-noise}
\Sigma_{k+1} 
= 
\big(A_{k} + B_{k} F_{k}\big) \Sigma_{k} \big(A_{k} + B_{k} F_{k}\big)\t + B_{k} V_{k} B_{k}\t, 
\end{equation}
where $F_{k} \in \R^{p \times n}$ and $V_{k} \in \R^{p \times p}$, $V_{k} \succeq 0$.

Let $\Phi_{A}(k, i)$ be the \emph{state transition matrix} of $A_{s}$ from time $i$ to $k$. 
That is, $\Phi_{A}(k, k) \triangleq I_{n}$ and, for $i < k$, $\Phi_{A}(k, i) \triangleq A_{k-1} A_{k-2} \cdots A_{i}$. 
Let $G(k, i)$ be the \emph{reachability Gramian} of the matrix pair $(A_{s}, B_{s})$ from time $i$ to $k$. 
That is, 
\begin{align*} 
&G(k, i) 
\\
&\triangleq 
\begin{cases}
0_{n \times n}, & \hspace{-2mm} i = k, 
\\* 
\sum_{j=i}^{k-1} \Phi_{A}(k, j+1) B_{j} B_{j}\t \Phi_{A}(k, j+1)\t, & \hspace{-2mm} i < k.
\end{cases}
\end{align*}

\begin{theorem} \label{thm:reach-no-noise}
When $D_{i} \equiv 0$ and the system~\eqref{sys:disc-time} starts from an initial state covariance $\Sigma_{0} \succeq 0$, the set of all reachable state covariances at time $k = 1, 2, \dots$ is
\begin{align*}
&\mathcal{R}_{k, 0} \triangleq \big\{\Sigma_{k} \succeq 0 : 
\\
&\qquad
P_{k, 0} \Sigma_{k} P_{k, 0} = 
P_{k, 0} \Phi_{A}(k, 0) \Sigma_{0} \Phi_{A}(k, 0)\t P_{k, 0} \big\}, 
\end{align*}
where $P_{k, 0}$ is the orthogonal projection onto the orthogonal complement of $\range G(k, 0)$. 
\end{theorem}

\noindent {\bf Proof:} 
We show Theorem~\ref{thm:reach-no-noise} by induction on $k$. 
When $k = 1$, since $\range G(1, 0) = \range B_{0}$, it follows from \eqref{eqn:stt-covar-no-noise} and Lemma~\ref{lem:sets} in the Appendix that $\Sigma_{1}$ is a positive semi-definite matrix with $P_{1, 0} \Sigma_{1} P_{1, 0} = P_{1, 0} A_{0} \Sigma_{0} A_{0}\t P_{1, 0}$. 
Hence, Theorem~\ref{thm:reach-no-noise} holds for $k = 1$.

Now, assume that Theorem~\ref{thm:reach-no-noise} holds for $k = s$. 
Then, it follows from Lemma~\ref{lem:sets} in the Appendix that $\bar{\Sigma}_{s} = \Phi_{A}(s, 0) \Sigma_{0} \Phi_{A}(s, 0)\t + G(s, 0) \bar{X} + \bar{X}\t G(s, 0)$ is reachable at time $s$, where $\bar{X} \in \R^{n \times n}$. 
It follows from \eqref{eqn:stt-covar-no-noise} and Lemma~\ref{lem:sets} that $\bar{\Sigma}_{s+1} = A_{s} \bar{\Sigma}_{s} A_{s}\t + B_{s} \bar{Y} + \bar{Y}\t B_{s}\t$ is reachable at time $s+1$, where $\bar{Y} \in \R^{p \times n}$. 
Hence, $\bar{\Sigma}_{s+1} = \Phi_{A}(s+1, 0) \Sigma_{0} \Phi_{A}(s+1, 0)\t + A_{s} G(s, 0) \bar{X} A_{s}\t + A_{s} \bar{X}\t G(s, 0) A_{s}\t + B_{s} \bar{Y} + \bar{Y}\t B_{s}\t$. 
Since $\range G(s+1, 0) = \range A_{s} G(s, 0) + \range B_{s}$, we have $P_{s+1, 0} \bar{\Sigma}_{s+1} P_{s+1, 0} = P_{s+1, 0} \Phi_{A}(s+1, 0) \Sigma_{0} \Phi_{A}(s+1, 0)\t P_{s+1, 0}$. 
That is, $\bar{\Sigma}_{s+1} \in \mathcal{R}_{s+1, 0}$. 
We have thus shown that all reachable state covariances at time $s+1$ belong to the set $\mathcal{R}_{s+1, 0}$.

Next, we show the other direction, i.e., each element of $\mathcal{R}_{s+1, 0}$ is a reachable state covariance at time $s+1$. 
Let $\bar{\Sigma}_{s+1} \in \mathcal{R}_{s+1, 0}$. 
Then, it follows from Lemma~\ref{lem:sets} in the Appendix that
\begin{multline*}
\bar{\Sigma}_{s+1}
= 
\big[\Phi_{A}(s+1, 0) ~~ A_{s} G(s, 0) ~~ B_{s}\big]
\\
\times
\begin{bmatrix}
\Sigma_{0} & \star_{12} & \star_{13} \\
\star_{21} & \star_{22} & \star_{23} \\
\star_{31} & \star_{32} & \star_{33}
\end{bmatrix}
\begin{bmatrix}
\Phi_{A}(s+1, 0)\t \\
G(s, 0) A_{s}\t \\
B_{s}\t
\end{bmatrix}, 
\end{multline*}
where the matrix in the middle of the right-hand side is a positive semi-definite matrix with $\Sigma_{0}$ on its top left block, 
and the $\star$ entries are irrelevant. 
In light of Lemma~\ref{lem:sets} in the Appendix
\begin{equation*}
\bar{\Sigma}_{s}
= 
\big[\Phi_{A}(s, 0) ~~ G(s, 0)\big]
\!
\begin{bmatrix}
\Sigma_{0} & \star_{12} \\
\star_{21} & \star_{22} 
\end{bmatrix}
\!
\begin{bmatrix}
\Phi_{A}(s, 0)\t \\
G(s, 0)
\end{bmatrix}
\in \mathcal{R}_{s, 0}, 
\end{equation*}
thus $\bar{\Sigma}_{s}$ is reachable by the induction assumption. 
Since $\bar{\Sigma}_{s+1} = A_{s} \bar{\Sigma}_{s} A_{s}\t + B_{s} \bar{Z} + \bar{Z}\t B_{s}\t \succeq 0$ for some $\bar{Z} \in \R^{p \times n}$, it follows from \eqref{eqn:stt-covar-no-noise} and Lemma~\ref{lem:sets} that $\bar{\Sigma}_{s+1}$ is reachable. 
Thus, Theorem~\ref{thm:reach-no-noise} holds for $k = s+1$. 
\hfill $\qed$

\begin{remark}
Let $\mu_{k} \triangleq \mathbb{E}[x_{k}] \in \R^{n}$ denote the mean value of the state $x_{k}$ of the system~\eqref{sys:disc-time}. 
Note that, at time $k = 1, 2, \dots$, the state mean $\mu_{k}$ is reachable from some initial state mean $\mu_{0}$ if and only if $\mu_{k} \in \Phi_{A}(k, 0) \mu_{0} + \range G(k, 0)$. 
Clearly, at time $k$, the set of all reachable mean values of the state of the system~\eqref{sys:disc-time} is $\{\mu_{k} \in \R^{n} : P_{k, 0} \mu_{k} = P_{k, 0} \Phi_{A}(k, 0) \mu_{0}\}$, which bears a close resemblance to the set $\mathcal{R}_{k, 0}$ of reachable state covariances in Theorem~\ref{thm:reach-no-noise}. 
In addition, if $\mu_{0} = 0$, the set of all reachable mean states of \eqref{sys:disc-time} at time $k$ is given by $\range G(k, 0)$, which is a linear subspace in $\R^{n}$. 
Likewise, if the initial state covariance $\Sigma_{0} = 0$ and the system~\eqref{sys:disc-time} has no additive noise, the set of all reachable state covariances of \eqref{sys:disc-time} at time $k$ is given by $\cone G(k, 0) \triangleq \{\Sigma \succeq 0 : \range \Sigma = \range G(k, 0)\}$, which is a convex subcone in the cone of $n \times n$ positive semi-definite matrices. 
However, we point out that when $\mu_{0} \neq 0$, the set of reachable mean states of \eqref{sys:disc-time} is an affine subspace, while when $\Sigma_{0} \neq 0$, the set of reachable state covariances of \eqref{sys:disc-time} is, in general, not an affine subcone. 
\end{remark}


\subsection{Reachability of the State Covariance With Noise}

Under the influence of additive noise, the state covariance of \eqref{sys:disc-time} propagates as in \eqref{eqn:stt-covar-fb}, that is, 
\begin{equation*} 
\Sigma_{k+1} 
\hspace{-0.3mm} = \hspace{-0.3mm} 
\big(A_{k} + B_{k} F_{k}\big) \Sigma_{k} \big(A_{k} + B_{k} F_{k}\big)\t \hspace{-0.1mm} + B_{k} V_{k} B_{k}\t + D_{k} D_{k}\t, 
\end{equation*}
where $F_{k} \in \R^{p \times n}$ and $V_{k} \in \R^{p \times p}$, $V_{k} \succeq 0$.

\begin{theorem} \label{thm:reach-noise}
When system~\eqref{sys:disc-time} starts from a given initial state covariance $\Sigma_{0} \succeq 0$ at time $0$, the set of all reachable state covariances at time $k = 1, 2, \dots$ is $\mathcal{R}_{k} \triangleq \sum_{i=0}^{k} \mathcal{R}_{k, i}$%
\footnote{The sum of the sets $\mathcal{S}_{0}, \mathcal{S}_{1}, \dots, \mathcal{S}_{k}$ is $\sum_{i=0}^{k} \mathcal{S}_{i} \triangleq \big\{\sum_{i=0}^{k} s_{i} : s_{i} \in \mathcal{S}_{i}, i = 0, 1, \dots, k\big\}$.}, 
where $\mathcal{R}_{k, 0}$ is defined in Theorem~\ref{thm:reach-no-noise} and, for $i = 1, 2, \dots, k$, 
\begin{align*}
&\mathcal{R}_{k, i} \triangleq 
\big\{\Sigma_{k, i} \succeq 0 : 
\\
&\hspace{3mm}
P_{k, i} \Sigma_{k, i} P_{k, i} = 
P_{k, i} \Phi_{A}(k, i) D_{i-1} D_{i-1}\t \Phi_{A}(k, i)\t P_{k, i} \big\}, 
\end{align*}
with $P_{k, i}$ being the orthogonal projection onto the orthogonal complement of $\range G(k, i)$. 
\end{theorem}

\begin{figure*}[b]
\hrulefill
\begin{align*}
&\bar{\Sigma}_{k} = 
\begin{bmatrix}
\Phi_{A}(k, 0) & \Phi_{A}(k, 1) & \cdots & A_{k-1} & I_{n} & \Phi_{A}(k, 1) B_{0} & \Phi_{A}(k, 2) B_{1} & \cdots & B_{k-1}
\end{bmatrix}
\\
&\hspace{3mm} \times \hspace{-1mm}
\begin{bmatrix}
\begin{array}{ccccc|cccc}
\Sigma_{0} &  &  &  &  & \star_{1, k+2} & \star_{1, k+3} & \cdots & \star_{1, 2k+1} \\
 & D_{0} D_{0}\t &  &  &  & \ostar_{2, k+2} & \star_{2, k+3} & \cdots & \star_{2, 2k+1} \\
 &  & \ddots &  &  & \vdots & \ddots & \ddots & \vdots \\
 &  &  & D_{k-2} D_{k-2}\t &  & \ostar_{k, k+2} & \ostar_{k, k+3} & \ddots & \star_{k, 2k+1} \\
 &  &  &  & D_{k-1} D_{k-1}\t & \ostar_{k+1, k+2} & \ostar_{k+1, k+3} & \cdots & \ostar_{k+1, 2k+1} \\
\hline
\star_{1, k+2}\t & \ostar_{2, k+2}\t & \cdots & \ostar_{k, k+2}\t & \ostar_{k+1, k+2}\t & \star_{k+2, k+2} & \star_{k+2, k+3} & \cdots & \star_{k+2, 2k+1} \\
\star_{1, k+3}\t & \star_{2, k+3}\t & \ddots & \ostar_{k, k+3}\t & \ostar_{k+1, k+3}\t & \star_{k+2, k+3}\t & \star_{k+3, k+3} & \cdots & \star_{k+3, 2k+1} \\
\vdots & \vdots & \ddots & \ddots & \vdots & \vdots & \vdots & \ddots & \vdots \\
\star_{1, 2k+1}\t & \star_{2, 2k+1}\t & \cdots & \star_{k, 2k+1}\t & \ostar_{k+1, 2k+1}\t & \star_{k+2, 2k+1}\t & \star_{k+3, 2k+1}\t & \cdots & \star_{2k+1, 2k+1}
\end{array}
\end{bmatrix}
\hspace{-1mm}
\begin{bmatrix}
\Phi_{A}(k, 0)\t \\
\Phi_{A}(k, 1)\t \\
\vdots \\
A_{k-1}\t \\
I_{n} \\
B_{0}\t \Phi_{A}(k, 1)\t \\
B_{1}\t \Phi_{A}(k, 2)\t \\
\vdots \\
B_{k-1}\t
\end{bmatrix} \hspace{-1mm} , 
\tag{6}
\end{align*}
\end{figure*}

\noindent {\bf Proof:} 
We show Theorem~\ref{thm:reach-noise} by induction on $k$. 
When $k = 1$, it follows from \eqref{eqn:stt-covar-fb} and Theorem~\ref{thm:reach-no-noise} that the set of reachable state covariances is $\mathcal{R}_{1, 0} + D_{0} D_{0}\t = \mathcal{R}_{1, 0} + \mathcal{R}_{1, 1} = \mathcal{R}_{1}$.

Assume that Theorem~\ref{thm:reach-noise} holds for $k = s$. 
That is, at time $s$, the set of all reachable state covariances is $\mathcal{R}_{s}$. 
It follows from \eqref{eqn:stt-covar-fb} that, at time $s+1$, the set of all reachable state covariances is $\big\{\Sigma_{s+1} \succeq 0 : \Sigma_{s+1} = (A_{s} + B_{s} F_{s}) \bar{\Sigma}_{s} (A_{s} + B_{s} F_{s})\t + B_{s} V_{s} B_{s}\t + D_{s} D_{s}\t,~\text{where}~F_{s} \in \R^{p \times n}, V_{s} \in \R^{p \times p}, V_{s} \succeq 0, \bar{\Sigma}_{s} \in \mathcal{R}_{s}\big\}$. 
Since $\mathcal{R}_{s} = \sum_{i=0}^{s} \mathcal{R}_{s, i}$, it follows from Lemma~\ref{lem:sepr} in the Appendix that the set of reachable state covariances at time $s+1$ is equal to $D_{s} D_{s}\t + \sum_{i=0}^{s} \mathcal{S}_{s+1, i}$, where $\mathcal{S}_{s+1, i} \triangleq \big\{\Sigma_{s+1, i} \succeq 0 : \Sigma_{s+1, i} = (A_{s} + B_{s} F_{s, i}) \bar{\Sigma}_{s, i} (A_{s} + B_{s} F_{s, i})\t + B_{s} V_{s, i} B_{s}\t, F_{s, i} \in \R^{p \times n}, V_{s, i} \in \R^{p \times p}, V_{s, i} \succeq 0, \bar{\Sigma}_{s, i} \in \mathcal{R}_{s, i}\big\}$. 
In light of Theorem~\ref{thm:reach-no-noise} and equation~\eqref{eqn:stt-covar-no-noise}, it is clear that, for all $i = 0, 1, \dots, s$, $\mathcal{S}_{s+1, i} = \mathcal{R}_{s+1, i}$. 
Since $\mathcal{R}_{s+1, s+1} = \{D_{s} D_{s}\t\}$, it follows that, at time $s+1$, the set of reachable state covariances is $\sum_{i=0}^{s+1} \mathcal{R}_{s+1, i} = \mathcal{R}_{s+1}$. 
Thus, Theorem~\ref{thm:reach-noise} holds for $k = s+1$, which completes the proof. 
\hfill $\qed$

\setcounter{equation}{6} 

\begin{remark}
Notice that the set $\mathcal{R}_{k, k}$ contains a single element, namely, $D_{k-1} D_{k-1}\t$. 
Thus, Theorem~\ref{thm:reach-noise} is consistent with the fact that $\Sigma_{k} \succeq D_{k-1} D_{k-1}\t$, which is reported in~\cite{goldshtein2017finite}. 
Intuitively, at each time $i = 1, 2, \dots, k$, we can treat $D_{i-1} D_{i-1}\t \succeq 0$ as the recurrent ``initial'' state covariance starting from time $i$ to time $k$. 
In this sense, $\Sigma_{0}$ can be regarded as $D_{-1} D_{-1}\t$. 
\end{remark}


Next, we give an alternative expression of the set $\mathcal{R}_{k}$ of the reachable state covariances at time $k$.

\begin{theorem} \label{thm:reach-noise-2}
The set $\mathcal{R}_{k}$ defined in Theorem~\ref{thm:reach-noise} can be written, equivalently, as $\mathcal{R}_{k} = \bigcap_{i=0}^{k} \mathcal{T}_{k, i}$, where
\begin{align*}
\mathcal{T}_{k, 0} \hspace{-0.5mm} &\triangleq \hspace{-0.5mm} 
\bigg\{ \hspace{-0.3mm} \Sigma_{k} \succeq 0 \hspace{-0.5mm} : \hspace{-0.5mm} 
P_{k, 0} \Sigma_{k} P_{k, 0} \hspace{-0.5mm} = \hspace{-0.5mm} P_{k, 0} \hspace{-0.5mm} \bigg[\Phi_{A}(k, 0) \Sigma_{0} \Phi_{A}(k, 0)\t 
\\*
&\hspace{14mm}
+ \sum_{j=1}^{k} \Phi_{A}(k, j) D_{j-1} D_{j-1}\t \Phi_{A}(k, j)\t\bigg] P_{k, 0}\bigg\}, 
\end{align*}
and, for all $i = 1, 2, \dots, k$, 
\begin{align*}
\mathcal{T}_{k, i} &\triangleq 
\bigg\{\Sigma_{k} \succeq 0 : 
P_{k, i} \Sigma_{k} P_{k, i} \succeq 
\\
&\hspace{8mm}
P_{k, i} \bigg[\sum_{j=i}^{k} \Phi_{A}(k, j) D_{j-1} D_{j-1}\t \Phi_{A}(k, j)\t\bigg] P_{k, i}\bigg\},
\end{align*}
with $P_{k, i}$ being the orthogonal projection onto the orthogonal complement of $\range G(k, i)$. 
\end{theorem}

\noindent {\bf Proof:} 
Let $\bar{\Sigma}_{k} \in \mathcal{R}_{k}$. 
Then, $\bar{\Sigma}_{k} = \sum_{i=0}^{k} \bar{\Sigma}_{k, i}$, where $\bar{\Sigma}_{k, i} \in \mathcal{R}_{k, i}$. 
Since, for $0 \leq i \leq j \leq k$, $\range G(k, j) \subseteq \range G(k, i)$, it follows that 
\begin{equation*}
P_{k, i} \bar{\Sigma}_{k, j} P_{k, i} = P_{k, i} \Phi_{A}(k, j) D_{j-1} D_{j-1}\t \Phi_{A}(k, j)\t P_{k, i}. 
\end{equation*}
Hence, for all $i = 0, 1, \dots, k$, $\bar{\Sigma}_{k} \in \mathcal{T}_{k, i}$. 
Since $\bar{\Sigma}_{k} \in \bigcap_{i=0}^{k} \mathcal{T}_{k, i}$, we have $\mathcal{R}_{k} \subseteq \bigcap_{i=0}^{k} \mathcal{T}_{k, i}$.

Next, we show the other direction. 
Let $\bar{\Sigma}_{k} \in \bigcap_{i=0}^{k} \mathcal{T}_{k, i}$. 
Since $\range G(k, 0) = \sum_{i=1}^{k} \range \Phi_{A}(k, i) B_{i-1}$, it follows that $P_{k, 0}$ is also the orthogonal projection onto $\big(\sum_{i=1}^{k} \range \Phi_{A}(k, i) B_{i-1}\big)^{\perp}$. 
Then, it follows from Lemma~\ref{lem:sets} in the Appendix and the fact $\bar{\Sigma}_{k} \in \mathcal{T}_{k, 0}$, that equation~(6) at the bottom of this page holds, 
where the matrix in the middle of the right-hand side of~(6) is a positive semi-definite matrix, denoted by $\Omega$, with its upper left block equal to $\blkdiag \big[\Sigma_{0}, D_{0} D_{0}\t, \dots, D_{k-1} D_{k-1}\t\big]$. 
In light of Proposition~\ref{prp:pos-semi-defn}, it follows from equation~(6) and the fact $\bar{\Sigma}_{k} \in \mathcal{T}_{k, 1}$, that 
$\ostar_{2, k+2} = \ostar_{3, k+2} = \cdots = \ostar_{k+1, k+2} = 0$. 
Similarly, for $i = 2, 3, \dots, k$, the fact $\bar{\Sigma}_{k} \in \mathcal{T}_{k, i}$ implies that $\ostar_{i+1, k+i+1} = \ostar_{i+2, k+i+1} = \cdots = \ostar_{k+1, k+i+1} = 0$ in addition to the previously known zero entries on the left of the same row(s) in the upper right block of $\Omega$. 
According to this partition, we denote the upper left (respectively, upper right, bottom left, bottom right) block of $\Omega$ by $\Omega_{UL}$ (respectively, $\Omega_{UR}$, $\Omega_{BL}$, $\Omega_{BR}$). 
Since $\Omega \succeq 0$, it follows from Proposition~\ref{prp:pos-semi-defn} that $\range \Omega_{UR} \subseteq \range \Omega_{UL}$ and $\Omega_{BR} \succeq \Omega_{BL} \Omega_{UL}^{+} \Omega_{UR}$. 
Since $\range \Omega_{UR} \subseteq \range \Omega_{UL}$, we have $\range [\star_{1, k+2} ~~ \star_{1, k+3} ~~ \cdots ~~ \star_{1, 2k+1}] \subseteq \range \Sigma_{0}$ and $\range [\star_{i+2, k+i+3} ~~ \star_{i+2, k+i+4} ~~ \cdots ~~ \star_{i+2, 2k+1}] \subseteq \range D_{i} D_{i}\t$ for $i = 0, 1, \dots, k-2$. 
Since $\Omega_{BR} \succeq \Omega_{BL} \Omega_{UL}^{+} \Omega_{UR}$, we have that 
\begin{align*}
\Omega_{BR} 
&\succeq \!
\begin{bmatrix}
\star_{1, k+2}\t \\
\star_{1, k+3}\t \\
\vdots \\
\star_{1, 2k+1}\t
\end{bmatrix}
\!
\Sigma_{0}^{+}
\big[\star_{1, k+2} ~~ \star_{1, k+3} ~~ \cdots ~~ \star_{1, 2k+1}\big]
\\
&\hspace{5mm}
+ \!
\begin{bmatrix}
\star_{2, k+3}\t \\
\vdots \\
\star_{2, 2k+1}\t
\end{bmatrix}
\!
(D_{0} D_{0}\t)^{+}
\big[\star_{2, k+3} ~~ \cdots ~~ \star_{2, 2k+1}\big]
\\
&\hspace{5mm}
+ 
\cdots 
+
\star_{k, 2k+1}\t (D_{k-2} D_{k-2}\t)^{+} \star_{k, 2k+1}. 
\end{align*}
Therefore, we can take $\bar{\Sigma}_{k, k} = D_{k-1} D_{k-1}\t$, 
\begin{align*}
&\bar{\Sigma}_{k, k-1} = 
\big[A_{k-1} ~~ B_{k-1}\big]
\\
&\hspace{2mm}
\times \hspace{-1mm}
\begin{bmatrix}
D_{k-2} D_{k-2}\t & \star_{k, 2k+1} \\
\star_{k, 2k+1}\t & \star_{k, 2k+1}\t (D_{k-2} D_{k-2}\t)^{+} \star_{k, 2k+1}
\end{bmatrix}
\hspace{-2mm}
\begin{bmatrix}
A_{k-1}\t \\
B_{k-1}\t
\end{bmatrix} \hspace{-1mm} , 
\\
&\bar{\Sigma}_{k, k-2} = 
\\
&\hspace{7mm}
\big[A_{k-1} A_{k-2} ~~ A_{k-1} B_{k-2} ~~ B_{k-1}\big]
\Omega_{k-2}
\hspace{-0.5mm}
\begin{bmatrix}
A_{k-2}\t A_{k-1}\t \\
B_{k-2}\t A_{k-1}\t \\
B_{k-1}\t
\end{bmatrix} \hspace{-1mm} , 
\end{align*}
where, 
\begin{align*}
&\Omega_{k-2} = 
\begin{bmatrix}
D_{k-3} D_{k-3}\t & \star_{k-1, 2k} & \star_{k-1, 2k+1} \\
\star_{k-1, 2k}\t & \star & \star \\
\star_{k-1, 2k+1}\t & \star & \star
\end{bmatrix}, 
\\
&\Omega_{k-2}/D_{k-3} D_{k-3}\t = 0, 
\end{align*}
and 
\begin{align*}
&\bar{\Sigma}_{k, 1} = 
\\
&\hspace{3mm}
\big[\Phi_{A}(k, 1) ~~ \Phi_{A}(k, 2) B_{1} ~~ \cdots ~~ B_{k-1}\big]
\Omega_{1}
\hspace{-0.5mm}
\begin{bmatrix}
\Phi_{A}(k, 1)\t \\
B_{1}\t \Phi_{A}(k, 2)\t \\
\vdots \\
B_{k-1}\t
\end{bmatrix} \hspace{-1mm} , 
\end{align*}
where, 
\begin{equation*}
\Omega_{1} \hspace{-0.5mm} = \hspace{-0.5mm} 
\begin{bmatrix}
D_{0} D_{0}\t & \star_{2, k+3} & \cdots & \star_{2, 2k+1} \\
\star_{2, k+3}\t & \star & \cdots & \star \\
\vdots & \vdots & \ddots & \vdots \\
\star_{2, 2k+1}\t & \star & \cdots & \star
\end{bmatrix} \hspace{-1mm} ,~~ 
\Omega_{1}/D_{0} D_{0}\t = 0. 
\end{equation*}
Finally, let 
\begin{align*}
&\bar{\Sigma}_{k, 0} = \bar{\Sigma}_{k} - \sum_{i=1}^{k} \bar{\Sigma}_{k, i} = 
\\
&\hspace{3mm}
\big[\Phi_{A}(k, 0) ~~ \Phi_{A}(k, 1) B_{0} ~~ \cdots ~~ B_{k-1}\big]
\Omega_{0}
\hspace{-0.5mm}
\begin{bmatrix}
\Phi_{A}(k, 0)\t \\
B_{0}\t \Phi_{A}(k, 1)\t \\
\vdots \\
B_{k-1}\t
\end{bmatrix} \hspace{-1mm} , 
\end{align*}
where, 
\begin{equation*}
\Omega_{0} \hspace{-0.5mm} = \hspace{-0.5mm} 
\begin{bmatrix}
\Sigma_{0} & \star_{1, k+2} & \cdots & \star_{1, 2k+1} \\
\star_{1, k+2}\t & \star & \cdots & \star \\
\vdots & \vdots & \ddots & \vdots \\
\star_{1, 2k+1}\t & \star & \cdots & \star
\end{bmatrix} \hspace{-1mm} ,~~ 
\Omega_{0}/\Sigma_{0} \succeq 0. 
\end{equation*}
In light of Proposition~\ref{prp:pos-semi-defn}, we have, for all $i = 0, 1, \dots, k-2$, $\Omega_{i} \succeq 0$. 
It follows from Lemma~\ref{lem:sets} in the Appendix that, for all $i = 0, 1, \dots, k$, $\bar{\Sigma}_{k, i} \in \mathcal{R}_{k, i}$. 
Thus, $\bar{\Sigma}_{k} \in \mathcal{R}_{k}$. 
Therefore, $\bigcap_{i=0}^{k} \mathcal{T}_{k, i} \subseteq \mathcal{R}_{k}$. 
\hfill $\qed$


\subsection{Controllability of the State Covariance}

In this subsection, necessary and sufficient conditions for the controllability of the state covariance of \eqref{sys:disc-time} are given.
These resemble the condition for the controllability of $(A_{k}, B_{k})$, equivalently, of the mean value of the state of \eqref{sys:disc-time}.

The following assumption is needed to derive one of the conditions for the controllability of the state covariance, which ensures that the deterministic system $x_{k+1} = A_{k} x_{k} + B_{k} u_{k}$, where $k = 0, 1, 2, \dots$, is time reversible.

\begin{assumption} \label{asm:A-inv}
For all $k = 0, 1, 2, \dots$, the matrix $A_{k}$ is invertible.
\end{assumption}

Let $\hat{G}(k, i)$ denote the \emph{controllability Gramian} of the matrix pair $(A_{s}, B_{s})$ from time $i$ to $k$. 
That is, $\hat{G}(k, i) \triangleq \Phi_{A}(i, k) G(k, i) \Phi_{A}(i, k)\t$, where $G(k, i)$ is the reachability Gramian defined above Theorem~\ref{thm:reach-no-noise}. 
Note that, under Assumption~\ref{asm:A-inv}, $\hat{G}(k, i) \succ 0$ if and only if $G(k, i) \succ 0$.

\begin{theorem} \label{thm:contr-noise}
The following statements are equivalent. 
\begin{enumerate}[leftmargin=*]

\item[i)] The state covariance of the system \eqref{sys:disc-time} is controllable from time $0$ to $k$. 

\item[ii)] $G(k, 0) \succ 0$ and $\range \Phi_{A}(k, i) D_{i-1} \subseteq \range G(k, i)$ for all $i = 1, 2, \dots, k$. 

\item[iii)] Under Assumption~\ref{asm:A-inv}, $\hat{G}(k, 0) \succ 0$ and, for all $i = 1, 2, \dots, k$, $\range D_{i-1} \subseteq \range \hat{G}(k, i)$. 

\end{enumerate}
\end{theorem}

\noindent {\bf Proof:} 
It follows from the definition of the controllability Gramian that $ii) \iff iii)$. 
Next, we show that $i) \iff ii)$. 
Let $\mathcal{S}^{+}$ denote the set of all $n \times n$ positive semi-definite matrices. 
In light of Theorem~\ref{thm:reach-noise}, the state covariance of \eqref{sys:disc-time} is controllable from time $0$ to $k$ if and only if $\mathcal{R}_{k} = \mathcal{S}^{+}$.

If $G(k, 0) \succ 0$, then, $P_{k, 0} = 0$. 
It follows that $\mathcal{R}_{k, 0} = \mathcal{S}^{+}$. 
If, for each $i = 1, 2, \dots, k$, $\range \Phi_{A}(k, i) D_{i-1} \subseteq \range G(k, i)$, then, we have $0_{n \times n} \in \mathcal{R}_{k, i} \subseteq \mathcal{S}^{+}$. 
It follows that $\mathcal{R}_{k} = \sum_{i=0}^{k} \mathcal{R}_{k, i} = \mathcal{S}^{+}$. 
Thus, $ii) \implies i)$.

Next, we show the other direction. 
It follows from \eqref{eqn:stt-covar-fb} that 
\begin{align} \label{eqn:stt-covar-k}
\Sigma_{k} &= \Phi_{A+BF}(k, 0) \Sigma_{0} \Phi_{A+BF}(k, 0)\t + \sum_{i = 1}^{k} \Phi_{A+BF}(k, i) 
\nonumber \\
&\hspace{3mm}
\times \!
\big(D_{i-1} D_{i-1}\t \! + \! B_{i-1} V_{i-1} B_{i-1}\t\big) \Phi_{A+BF}(k, i)\t, 
\end{align}
where $\Phi_{A+BF}(k, i)$ is the state transition matrix of $A_{s} + B_{s} F_{s}$ from time $i = 0, 1, \dots, k$ to time $k$. 
If $\mathcal{R}_{k} = \mathcal{S}^{+}$, then, clearly, for any $\Sigma_{0} \succ 0$, there exists $F_{i} \in \R^{p \times n}$ and $V_{i} \in \R^{p \times p}$, $V_{i} \succeq 0$, $i = 0, 1, \dots, k$, such that $\Sigma_{k} = 0_{n \times n}$. 
Let $V_{i} \equiv 0$. 
In view of \eqref{eqn:stt-covar-k}, it follows that there exists $F_{s}$ such that $\Phi_{A+BF}(k, 0) = 0$ and, for all $i = 1, 2, \dots, k$, $\Phi_{A+BF}(k, i) D_{i-1} = 0$. 
Since the deterministic system $x_{s+1} = (A_{s} + B_{s} F_{s}) x_{s}$, where $s = 0, 1, 2, \dots$, satisfies $x_{k} = \Phi_{A+BF}(k, 0) x_{0}$, it follows from $\Phi_{A+BF}(k, 0) = 0$ that the pair $(A_{s}, B_{s})$ is controllable from time $0$ to $k$. 
Thus, $G(k, 0) \succ 0$. 
Similarly, for each $i = 1, 2, \dots, k$, the fact that there exists $F_{s}$ such that $\Phi_{A+BF}(k, i) D_{i-1} = 0$ implies that $\range \Phi_{A}(k, i) D_{i-1} \subseteq \range G(k, i)$. 
Therefore, $i) \implies ii)$. 
\hfill $\qed$

\begin{remark}
Notice that in the second statement of Theorem~\ref{thm:contr-noise}, for $i = k$, we obtain that $\range \Phi_{A}(k, k) D_{k-1} = \range D_{k-1} \subseteq \range G(k, k) = 0$, i.e., $D_{k-1} = 0$. 
In other words, if the state covariance of the system \eqref{sys:disc-time} is controllable from time $0$ to $k$, we must have $D_{k-1} = 0$. 
Recall that the matrix pair $(A_{s}, B_{s})$ is controllable from time $0$ to $k$ if and only if $G(k, 0) \succ 0$, which is equivalent to $\hat{G}(k, 0) \succ 0$ under Assumption~\ref{asm:A-inv}. 
Moreover, in case $G(k, 0) \succeq 0$, for a given initial state $\bar{x}_{0} \in \R^{n}$, there exists a control that steers the state of the deterministic system $x_{s+1} = A_{s} x_{s} + B_{s} u_{s}$ from $x_{0} = \bar{x}_{0}$ to $x_{k} = 0$ if and only if $\Phi_{A}(k, 0) \bar{x}_{0} \in \range G(k, 0)$, which is equivalent to $\bar{x}_{0} \in \range \hat{G}(k, 0)$ under Assumption~\ref{asm:A-inv}. 
Clearly, at each $i = 1, 2, \dots, k$, if we treat $D_{i-1} D_{i-1}\t \succeq 0$ as the recurrent ``initial'' state covariance starting from time $i$ to time $k$, then, the conditions in Theorem~\ref{thm:contr-noise} resemble the conditions for the controllability of the matrix pair $(A_{s}, B_{s})$. 
\end{remark}

\begin{remark}
It is straightforward to check that 
\begin{equation*}
\range \Phi_{A}(k, i) D_{i-1} \subseteq \range G(k, i) 
\end{equation*}
if and only if 
\begin{equation*}
\Phi_{A}(k, i) D_{i-1} D_{i-1}\t \Phi_{A}(k, i)\t \in \cone G(k, i), 
\end{equation*}
where $\cone G(k, i) \triangleq \{\Sigma \succeq 0 : \range \Sigma = \range G(k, i)\}$ is a convex cone. 
\end{remark}


\begin{corollary} \label{cor:contr-no-noise}
Assume that there is no additive noise in the system~\eqref{sys:disc-time}, that is, $D_{i} \equiv 0$. 
Then, the state covariance of \eqref{sys:disc-time} is controllable from time $0$ to $k$ if and only if the matrix pair $(A_{s}, B_{s})$ is controllable from time $0$ to $k$. 
\end{corollary}

\begin{corollary} \label{cor:contr-all-noise}
Assume that in the system~\eqref{sys:disc-time}, the additive noise is non-degenerate at time $i = 0, 1, \dots, k-2$ and is zero at time $k-1$. 
That is, for all $i = 0, 1, \dots, k-2$, $\rank D_{i} = n$, and $D_{k-1} = 0$. 
Then, under Assumption~\ref{asm:A-inv}, the state covariance of \eqref{sys:disc-time} is controllable from time $0$ to $k$ if and only if $\rank B_{k-1} = n$. 
\end{corollary}


\section{Continuous-Time Case} \label{sec:continuous}

In this section, we derive results regarding the controllability of the continuous-time covariance equation. 
We first show that it suffices to consider the class of state-feedback controls for the system~\eqref{sys:cont-time}. 
Then, we give an upper bound on the set of all reachable state covariances. 
Finally, we derive a necessary and sufficient condition for the controllability of the state covariance of~\eqref{sys:cont-time} on a finite horizon.

\subsection{State Feedback Controls}

Let $\Sigma(t)$ denote the covariance of the state $x(t)$ at time $t$. 
It is clear that the state covariance of \eqref{sys:cont-time} satisfies 
\begin{multline} \label{ode:stt-covar}
\dot{\Sigma}(t) = A(t) \Sigma(t) + \Sigma(t) A(t)\t 
\\*
+ B(t) \Sigma^{ux}(t) + \Sigma^{ux}(t)\t B(t)\t + D(t) D(t)\t, 
\end{multline}
where $\Sigma^{ux}(t)$ is the cross-covariance of $u(t)$ and $x(t)$. 
Note that $\Sigma^{ux}(t)$ can be treated as the ``control'' in
equation \eqref{ode:stt-covar}. 
Note also that, in contrary to the discrete-time case, there is no $\Sigma^{uu}$ term in~\eqref{ode:stt-covar}.
It follows from Proposition~\ref{prp:pos-semi-defn} that, 
for each $t \in [0, T]$, there exists a deterministic matrix $F(t) \in \R^{p \times n}$, which may depend on the distributions of $u(t)$ and $x(t)$, such that $\Sigma^{ux}(t) = F(t) \Sigma(t)$. 
%
Hence, \eqref{ode:stt-covar} can be written as 
\begin{multline} \label{ode:stt-covar-fb}
\dot{\Sigma}(t) = \big[A(t) + B(t) F(t)\big] \Sigma(t) + \Sigma(t) \big[A(t) + B(t) F(t)\big]\t 
\\*
+ D(t) D(t)\t. 
\end{multline}
Thus, we have the following result.

\begin{proposition} \label{prp:stt-fb}
At time $T > 0$, if a terminal state covariance $\bar{\Sigma}_{T}$ of the continuous-time system~\eqref{sys:cont-time} is reachable from a given initial state covariance $\Sigma(0) = \Sigma_{0}$, then, $\bar{\Sigma}_{T}$ is also reachable via a state feedback control $u(t) = F(t) x(t)$, 
for some $F(t) \in \R^{p \times n}$, $t \in [0, T]$. 
\end{proposition}


\subsection{Upper Bound on Reachable State Covariances}

Let $\Phi_{A}(T, t)$ denote the state transition matrix from time $t$ to $T$ of the matrix $A(\tau)$. 
Let 
\begin{equation} \label{def:G-cont}
G(T, t) \triangleq \int_{t}^{T} \Phi_{A}(T, \tau) B(\tau) B(\tau)\t \Phi_{A}(T, \tau)\t \, \dd \tau,
\end{equation}
be the \emph{reachability Gramian} of the matrix pair $\big(A(\tau), B(\tau)\big)$ from time $t$ to $T$. 
It follows immediately from \eqref{ode:stt-covar-fb} that 
\begin{align} \label{eqn:stt-covar-fb-cont}
&\Sigma(T) = \Phi_{A+BF}(T, 0) \Sigma(0) \Phi_{A+BF}(T, 0)\t 
\nonumber \\*
&\hspace{6mm}
+ \int_{0}^{T} \Phi_{A+BF}(T, \tau) D(\tau) D(\tau)\t \Phi_{A+BF}(T, \tau)\t \dd \tau.
\end{align}

In order to bound the set of reachable state covariances of \eqref{sys:cont-time}, it suffices to bound $\Phi_{A+BF}(T, t)$. 
The following result follows immediately from Lemma~\ref{lem:bd} in the Appendix.

\begin{theorem} \label{thm:reach-noise-bd}
Let $\Sigma_{0} \succeq 0$ at time $0$. 
The set of reachable state covariances at time $T > 0$, denoted by $\mathcal{R}_{T}$, satisfies 
$\mathcal{R}_{T} \subseteq \mathcal{R}_{T}^{+}$, where, 
\begin{align*}
&\mathcal{R}_{T}^{+} \triangleq 
\big\{\Sigma_{T} \succeq 0 : \Sigma_{T} = 
\\*
&\hspace{1.5mm}
\big[\Phi_{A}(T, 0) \hspace{-0.5mm} + \hspace{-0.5mm} G(T, 0) K(0)\big] \Sigma_{0} \big[\Phi_{A}(T, 0) \hspace{-0.5mm} + \hspace{-0.5mm} G(T, 0) K(0)\big]\t
\\*
&\hspace{2mm}
+ \int_{0}^{T} \big[\Phi_{A}(T, \tau) + G(T, \tau) K(\tau)\big] D(\tau) D(\tau)\t 
\\*
&\hspace{2mm} \times 
\big[\Phi_{A}(T, \tau) + G(T, \tau) K(\tau)\big]\t \dd \tau\big\},
\end{align*}
where  $K(t) \in \R^{n \times n}$ is such that 
$\Phi_{A}(T, t) + G(T, t) K(t)$ is invertible 
for all $t \in [0, T]$.

\end{theorem}


\subsection{Controllability of the State Covariance}

In this section, we provide necessary and sufficient conditions for the controllability of the state covariance of \eqref{sys:cont-time}. 
To ths end, let $\hat{G}(T, t)$ denote the \emph{controllability Gramian} of the matrix pair $\big(A(\tau), B(\tau)\big)$ from time $t$ to $T$. 
That is, $\hat{G}(T, t) \triangleq \Phi_{A}(t, T) G(T, t) \Phi_{A}(t, T)\t$, where $G(T, t)$ is the reachability Gramian given by \eqref{def:G-cont}. 
Clearly, $\hat{G}(T, t) \succ 0$ if and only if $G(T, t) \succ 0$.

It is well known that the Riccati differential equation 
\begin{equation} \label{ode:pi-1}
\dot{\Pi} = - A(t)\t \Pi - \Pi A(t) + \Pi B(t) B(t)\t \Pi 
\end{equation}
may have ``finite escape'' times and therefore the solution may not exist on the whole interval $[0, T]$. 
Below, we provide a necessary and sufficient condition for \eqref{ode:pi-1} to admit a solution on $[0, T]$, that will be used to prove the main result of this section, and which may be of independent interest. 
It is worth pointing out that, unlike the condition in \cite[Lemma~2]{liu2024add}, the condition below does not require $\big(A(t), B(t)\big)$ to be totally controllable or completely controllable, thus it is more general than the result of~\cite{liu2024add}.

\begin{theorem} \label{thm:pi-exist}
The following statements are equivalent. 
\begin{enumerate}[leftmargin=*]

\item[i)] The Riccati differential equation \eqref{ode:pi-1} with initial condition $\Pi(0)$ admits a unique solution $\Pi(t)$ on $[0, T]$. 

\item[ii)] For all $t \in [0, T]$, the eigenvalues of $I_{n} - \hat{G}(t, 0) \Pi(0)$ are all positive. 

\item[iii)] The eigenvalues of $I_{n} - \hat{G}(T, 0) \Pi(0)$ are all positive. 

\item[iv)] For all $t \in [0, T]$, $\hat{G}(t, 0)^{\frac{1}{2}} \Pi(0) \hat{G}(t, 0)^{\frac{1}{2}} \prec I_{n}$. 

\item[v)] $\hat{G}(T, 0)^{\frac{1}{2}} \Pi(0) \hat{G}(T, 0)^{\frac{1}{2}} \prec I_{n}$. 

\end{enumerate}
Furthermore, the solution of the Riccati equation is given by
\begin{equation} \label{sol:pi}
\Pi(t) = \Phi_{A}(0, t)\t \Pi(0) \big(I_{n} - \hat{G}(t, 0) \Pi(0)\big)^{-1} \Phi_{A}(0, t). 
\end{equation}
\end{theorem}

The proof of Theorem~\ref{thm:pi-exist} is given in the Appendix. 
Next, we present the main result of this section on the controllability of the state covariance of \eqref{sys:cont-time}.

\begin{theorem} \label{thm:contr-noise-cont}
The following statements are equivalent. 
\begin{enumerate}[leftmargin=*]

\item[i)] The state covariance of the system \eqref{sys:cont-time} is controllable on the time interval $[0, T]$. 

\item[ii)] $G(T, 0) \succ 0$ and $\range \Phi_{A}(T, t) D(t) \subseteq \range G(T, t)$ for all $t \in [0, T)$. 

\item[iii)] $\hat{G}(T, 0) \succ 0$ and $\range D(t) \subseteq \range \hat{G}(T, t)$ for all $t \in [0, T)$. 

\end{enumerate}
\end{theorem}

\noindent {\bf Proof:} 
If the state covariance of \eqref{sys:cont-time} is controllable on $[0, T]$, then, for any $\bar{\Sigma}_{0} \succ 0$ and any $\bar{\Sigma}_{T} \succ 0$ arbitrarily close to zero, there exists $u(t) = F(t) x(t)$ that steers the system from $\Sigma(0) = \bar{\Sigma}_{0}$ to $\Sigma(T) = \bar{\Sigma}_{T}$. 
In view of \eqref{eqn:stt-covar-fb-cont}, it follows that $\Phi_{A+BF}(T, 0) \bar{\Sigma}_{0}$ is arbitrarily close to zero and, for all $t \in [0, T)$, $\Phi_{A+BF}(T, t) D(t)$ is arbitrarily close to zero. 
If $\range G(T, 0) \neq \R^{n}$, one can select $x_{0}$ such that $\Phi_{A}(T, 0) x_{0} \notin \range G(T, 0)$. 
Then, starting from the initial state $x(0) = x_{0}$, the set of all reachable states of the corresponding deterministic system $\dot{x}(t) = A(t) x(t) + B(t) u(t)$ at time $T$ is the affine subspace $\Phi_{A}(T, 0) x_{0} + \range G(T, 0)$. 
Since this affine subspace does not contain zero and $x_{0} \in \range \bar{\Sigma}_{0}$, it contradicts the fact that, for some $x_{T}$ arbitrarily close to zero, there exists $u(t) = F(t) x(t)$ that steers the deterministic system from $x(0) = x_{0}$ to $x(T) = x_{T} = \Phi_{A+BF}(T, 0) x_{0}$. 
Hence, $G(T, 0) \succ 0$. 
Similarly, for a given $t \in [0, T)$, the fact that there exists $F(t)$ such that $\Phi_{A+BF}(T, t) D(t)$ is arbitrarily close to zero implies that $\range \Phi_{A}(T, t) D(t) \subseteq \range G(T, t)$. 
Thus, $i) \implies ii)$.


Clearly, from the definitions of the reachability and controllability Gramians it follows that $ii) \iff iii)$.


To show $iii) \implies i)$, 
we first show that if $iii)$ holds, then, for any given $\bar{\Sigma}_{0}, \bar{\Sigma}_{T} \succ 0$, there exists an optimal control that steers the state covariance of the system~\eqref{sys:cont-time} from $\Sigma(0) = \bar{\Sigma}_{0}$ to $\Sigma(T) = \bar{\Sigma}_{T}$ with the minimum expected control energy. 
It is first shown in \cite{chen2016II} that such an optimal control is $u(t) = - B(t)\t \Pi(t) x(t)$, provided that $\Pi(t)$ solves the following coupled matrix differential equations 
\begin{subequations} \label{ode:coupled}
\begin{align}
&\dot{\Pi} = - A(t)\t \Pi - \Pi A(t) + \Pi B(t) B(t)\t \Pi,
\label{ode:pi} \\
&\dot{\Sigma} = \big(A(t) - B(t) B(t)\t \Pi(t)\big)\Sigma 
\nonumber \\*
&\hspace{10mm}
+ \Sigma\big(A(t)\t - \Pi(t) B(t) B(t)\t\big) + D(t) D(t)\t,
\label{ode:sigma} \\
&\Sigma(0) = \bar{\Sigma}_{0}, \qquad \Sigma(T) = \bar{\Sigma}_{T}. 
\label{bdr:sigma} 
\end{align}
\end{subequations}
Thus, it suffices to show that if $iii)$ holds, then, for given $\bar{\Sigma}_{0}, \bar{\Sigma}_{T} \succ 0$, the coupled equations \eqref{ode:coupled} have a solution. 
In light of Theorem~\ref{thm:pi-exist}, 
the Riccati differential equation \eqref{ode:pi} with initial condition $\Pi(0)$ admits the unique solution $\Pi(t) = \Phi_{A}(0, t)\t \Pi(0) \big(I_{n} - \hat{G}(t, 0) \Pi(0)\big)^{-1} \Phi_{A}(0, t)$ on $[0, T]$ if and only if the eigenvalues of $I_{n} - \hat{G}(T, 0) \Pi(0)$ are all positive. 
Since $\hat{G}(T, 0) \succ 0$, this condition reduces to $\Pi_{0} \prec \hat{G}(T, 0)^{-1}$.


Next, we borrow some results from \cite{liu2024add} to show that, if $iii)$ holds, then, for given $\bar{\Sigma}_{0}, \bar{\Sigma}_{T} \succ 0$, the coupled matrix differential equations \eqref{ode:coupled} admit a unique solution. 
In view of \cite[Lemma~3]{liu2024add}, let the state transition matrix of $A(\tau) - B(\tau) B(\tau)\t \Pi(\tau)$ from time $s$ to $t$ be denoted by
\begin{equation} \label{def:phi-pi}
\Phi_{\Pi}(t, s) \triangleq \Phi_{A-BB\t\Pi}(t, s) = \Phi_{A}(t, s) \big(I_{n} - \hat{G}(t, s) \Pi(s)\big). 
\end{equation}

Define the sets 
$\mathcal{P} \triangleq 
\big\lbrace 
\Pi_{0} \in \R^{n \times n} : \Pi_{0} \prec \hat{G}(T, 0)^{-1}
\big\rbrace$ 
and 
$\mathcal{M} \triangleq 
\big\lbrace 
\Sigma_{T} \in \R^{n \times n} : \Sigma_{T} \succ 0
\big\rbrace$. 
As in \cite[Section~V.C]{liu2024add}, given an initial state covariance $\bar{\Sigma}_{0} \succ 0$ of the system~\eqref{sys:cont-time}, we can define the map $f: \mathcal{P} \to \mathcal{M}$, according to 
\begin{align} \label{map:f}
f(\Pi_{0}) 
&= \Phi_{A}(T, 0) \big(I_{n} - \hat{G}(T, 0) \Pi_{0}\big) \bigg[ \bar{\Sigma}_{0} 
\nonumber \\*
&\hspace{3mm}
+ \int_{0}^{T} \big(I_{n} - \hat{G}(\tau, 0) \Pi_{0}\big)^{-1} \Phi_{A}(0, \tau) D(\tau) D(\tau)\t 
\nonumber \\*
&\hspace{15mm}
\times 
\Phi_{A}(0, \tau)\t 
\big(I_{n} - \Pi_{0} \hat{G}(\tau, 0)\big)^{-1} \dd \tau \bigg] 
\nonumber \\*
&\hspace{20mm}
\times 
\big(I_{n} - \Pi_{0} \hat{G}(T, 0)\big) \Phi_{A}(T, 0)\t, 
\end{align}
where, for notational simplicity, we write $\Pi_{0}$ for $\Pi(0)$.

Let $\vect(M)$ denote the vectorization of the matrix $M$. 
Define the vectorized map $\bar{f}$ such that $\bar{f}\big(\vect(\Pi_{0})\big) = \vect\big(f(\Pi_{0})\big)$, where the domain and codomain of $\bar{f}$ are the vectorized $\mathcal{P}$ and $\mathcal{M}$, respectively. 
Following the same procedure as in the proof of \cite[Lemma~4]{liu2024add}, it can be shown that the Jacobian $J_{\bar{f}}$ of $\bar{f}$ at $\vect(\Pi_{0})$ is 
\begin{align*} 
&J_{\bar{f}} \big(\vect(\Pi_{0})\big) 
= - \Phi_{\Pi}(T, 0) \otimes \Phi_{\Pi}(T, 0) \bigg[ \bar{\Sigma}_{0} \otimes Q(T) 
\\* 
&\hspace{8mm}
+ Q(T) \otimes \bar{\Sigma}_{0} 
+ \int_{0}^{T} U(\tau) \otimes \big(Q(T) - Q(\tau)\big) 
\\* 
&\hspace{25mm}
+ \big(Q(T) - Q(\tau)\big) \otimes U(\tau) \, \dd \tau \bigg],
\end{align*}
where $\otimes$ denotes the Kronecker product, 
for $t \in [0, T]$, $Q(t) \triangleq \big(I_{n} - \hat{G}(t, 0) \Pi_{0}\big)^{-1} \hat{G}(t, 0)$, 
and 
\begin{multline*} 
U(t) \triangleq \big(I_{n} - \hat{G}(t, 0) \Pi_{0}\big)^{-1} \Phi_{A}(0, t) D(t) D(t)\t 
\\*
\times 
\Phi_{A}(0, t)\t \big(I_{n} - \Pi_{0} \hat{G}(t, 0)\big)^{-1} \succeq 0.
\end{multline*}
Since $\hat{G}(T, 0) \succ 0$ and $\Pi_{0} \prec \hat{G}(T, 0)^{-1}$, we have $Q(T) \succ 0$. 
From \cite[Lemma~13]{liu2024disc} it follows that, for all $\Pi_{0} \in \mathcal{P}$ and $t \in [0, T]$, we have $Q(T) \succeq Q(t) \succeq 0$. 
Using the same technique as in \cite[Lemma~5]{liu2024add}, it is not difficult to show that the Jacobian $J_{\bar{f}} \big(\vect(\Pi_{0})\big)$ is nonsingular at each $\vect(\Pi_{0})$ with $\Pi_{0} \in \mathcal{P}$.


Finally, we show that for any given $\bar{\Sigma}_{0} \succ 0$, the map $f$ given by \eqref{map:f} is a homeomorphism. 
Therefore, for given $\bar{\Sigma}_{0}, \bar{\Sigma}_{T} \succ 0$, the coupled matrix differential equations \eqref{ode:coupled} admit a unique solution. 
In order to show that $f$ is a homeomorphism, we need to check the conditions in Hadamard's global inverse function theorem~\cite{krantz2002implicit}. 
First, $\mathcal{P}$ and $\mathcal{M}$ are smooth, connected manifolds of the same dimension. 
Second, $J_{\bar{f}} \big(\vect(\Pi_{0})\big)$ is continuous and nonsingular at each $\vect(\Pi_{0})$ for which $\Pi_{0} \in \mathcal{P}$. 
Third, since $\mathcal{M}$ is convex, it is simply connected. 
Next, we show that $f$ is proper. 
Clearly, the boundedness of $f(\Pi_{0})$ implies the boundedness of $\Phi_{A}(T, 0) \big(I_{n} - \hat{G}(T, 0) \Pi_{0}\big) \bar{\Sigma}_{0} \big(I_{n} - \Pi_{0} \hat{G}(T, 0)\big) \Phi_{A}(T, 0)\t$. 
Since $\Phi_{A}(T, 0)$, $\hat{G}(T, 0)$, and $\bar{\Sigma}_{0}$ are all invertible, it follows that $\Pi_{0}$ is bounded as well. 
Finally, we need to show that if $\mathcal{S} \subset \mathcal{M}$ is closed and bounded in $\R^{n \times n}$, then, the inverse image $f^{-1}(\mathcal{S}) \subset \mathcal{P}$ is closed in $\R^{n \times n}$. 
We claim that $f^{-1}(\mathcal{S})$ is bounded away from the boundary $\partial \mathcal{P}$ of $\mathcal{P}$. 
It then follows from the continuity of $f$ that $f^{-1}(\mathcal{S})$ is closed in $\R^{n \times n}$. 
In light of Hadamard's global inverse function theorem, $f$ is a homeomorphism.

Next, we show that $f^{-1}(\mathcal{S})$ is bounded away from $\partial \mathcal{P}$. 
Note that it suffices to show that $f$ maps $\partial \mathcal{P}$ to $\partial \mathcal{M}$ in the limit. 
In other words, we will show that, as $\Pi_{0}$ approaches $\partial \mathcal{P}$, then, $f(\Pi_{0})$ approaches $\partial \mathcal{M}$. 
If this statement holds, then it follows that $f^{-1}(\mathcal{S})$ is bounded away from $\partial \mathcal{P}$, since $\mathcal{S}$ is bounded away from $\partial \mathcal{M}$. 
The first case of $\Pi_{0}$ approaching $\partial \mathcal{P}$ is that the smallest eigenvalue of $\Pi_{0}$ goes to $-\infty$. 
Since $\hat{G}(T, 0) \succ 0$, it follows from Lemma~\ref{lem:spec} that one eigenvalue of $I_{n} - \hat{G}(T, 0) \Pi_{0}$ goes to $+\infty$. 
Since $\Phi_{A}(T, 0)$ is invertible and $\bar{\Sigma}_{0} \succ 0$, one eigenvalue of $\Phi_{A}(T, 0) \big(I_{n} - \hat{G}(T, 0) \Pi_{0}\big) \bar{\Sigma}_{0} \big(I_{n} - \Pi_{0} \hat{G}(T, 0)\big) \Phi_{A}(T, 0)\t$ goes to $+\infty$. 
It follows that the largest eigenvalue of $f(\Pi_{0})$ goes to $+\infty$. 
The second case of $\Pi_{0}$ approaching $\partial \mathcal{P}$ is that the smallest eigenvalue of $\hat{G}(T, 0)^{-1} - \Pi_{0}$ goes to zero. 
Let $\bar{P} \in \partial \mathcal{P}$ such that $\bar{P} \preceq \hat{G}(T, 0)^{-1}$ and the smallest eigenvalue of $\hat{G}(T, 0)^{-1} - \bar{P}$ is zero. 
Next, we show that, as $\Pi_{0} \to \bar{P}$, then the smallest eigenvalue of $f(\Pi_{0})$ goes to zero.

It follows from \eqref{ode:sigma} and \eqref{def:phi-pi} that 
\begin{align} \label{eqn:phi-f-phi}
&\Phi_{A}(0, T) f(\Pi_{0}) \Phi_{A}(0, T)\t 
= 
\nonumber \\
&\hspace{10mm}
\big(I_{n} - \hat{G}(T, 0) \Pi_{0}\big) \bar{\Sigma}_{0} \big(I_{n} - \Pi_{0} \hat{G}(T, 0)\big) 
\nonumber \\* 
&\hspace{15mm}
+ \int_{0}^{T} \Phi_{A}(0, \tau) \big(I_{n} - \hat{G}(T, \tau) \Pi_{\tau}\big) D(\tau) D(\tau)\t 
\nonumber \\* 
&\hspace{15mm}
\times \big(I_{n} - \Pi_{\tau} \hat{G}(T, \tau)\big) \Phi_{A}(0, \tau)\t \, \dd \tau, 
\end{align}
where, for $t \in [0, T]$, $\Pi_{t} = \Pi(t)$ is the solution to the Riccati differential equation \eqref{ode:pi} starting from $\Pi(0) = \Pi_{0}$. 
Since $\Phi_{A}(0, T)$ is invertible and $\bar{\Sigma}_{0} \succ 0$, it follows from the above equation that if the smallest eigenvalue of $f(\Pi_{0})$ goes to zero, then, one eigenvalue of $I_{n} - \hat{G}(T, 0) \Pi_{0}$ goes to zero. 
In light of Lemma~\ref{lem:spec} in the Appendix, only the positive eigenvalues of $\Pi_{0}$ can possibly render $I_{n} - \hat{G}(T, 0) \Pi_{0}$ singular. 
Hence, it suffices to consider the case when $\Pi_{0}, \bar{P} \succ 0$. 
Since $0 \prec \bar{P} \preceq \hat{G}(T, 0)^{-1}$, we can write $\bar{P} = \big(\hat{G}(T, 0) + \Gamma \Gamma\t\big)^{-1}$, where $\Gamma \in \R^{n \times \ell}$, $\Gamma \Gamma\t \succeq 0$. 
In view of the Woodbury matrix identity~\cite{horn2012matrix}, we have $\big(\hat{G}(T, 0) + \Gamma \Gamma\t\big)^{-1} = \hat{G}(T, 0)^{-1} - \hat{G}(T, 0)^{-1} \Gamma \big(I_{\ell} + \Gamma\t \hat{G}(T, 0)^{-1} \Gamma\big)^{-1} \Gamma\t \hat{G}(T, 0)^{-1}$. 
Thus, as $\Pi_{0} \to \bar{P}$, then, $I_{n} - \hat{G}(T, 0) \Pi_{0} \to I_{n} - \hat{G}(T, 0) \big(\hat{G}(T, 0) + \Gamma \Gamma\t\big)^{-1} = \Gamma \big(I_{\ell} + \Gamma\t \hat{G}(T, 0)^{-1} \Gamma\big)^{-1} \Gamma\t \hat{G}(T, 0)^{-1}$. 
Since $\Gamma \Gamma\t \succeq 0$ is singular, its left null space is nonempty. 
Let $z \in \ker (\Gamma \Gamma\t)$ be a nonzero vector, then, $z\t \Gamma = 0$. 
It follows that, as $\Pi_{0} \to \bar{P}$, then, $I_{n} - \hat{G}(T, 0) \Pi_{0}$ approaches some matrix for which $z$ is in its left null space. 
Next, we show that as $\Pi_{0} \to \bar{P}$, then, for each $t \in [0, T]$, $\Phi_{A}(0, t) \big(I_{n} - \hat{G}(T, t) \Pi_{t}\big) D(t)$ approaches some matrix for which $z$ is in its left null space as well.

Let $t \in [0, T]$ be fixed. 
Since $\range D(t) \subseteq \range \hat{G}(T, t)$ and $\Phi_{A}(0, t)\t$ is invertible, it suffices to consider the left null space of $\Phi_{A}(0, t) \big(I_{n} - \hat{G}(T, t) \Pi_{t}\big) \hat{G}(T, t) \Phi_{A}(0, t)\t$. 
Let $W \triangleq \Phi_{A}(0, t) \hat{G}(T, t) \Phi_{A}(0, t)\t = \hat{G}(T, 0) - \hat{G}(t, 0) \succeq 0$. 
From the equation $\Pi_{t} = \Phi_{A}(0, t)\t \Pi_{0} \big(I_{n} - \hat{G}(t, 0) \Pi_{0}\big)^{-1} \Phi_{A}(0, t)$, it follows that as $\Pi_{0} \to \bar{P}$, then, $\Pi_{t} \to \Phi_{A}(0, t)\t \big(\bar{P}^{-1} - \hat{G}(t, 0)\big)^{-1} \Phi_{A}(0, t) = \Phi_{A}(0, t)\t (W + \Gamma \Gamma\t)^{-1} \Phi_{A}(0, t)$.\footnote{Since $W \succeq 0$ and $\Gamma \Gamma\t \succeq 0$, $W + \Gamma \Gamma\t \succeq 0$ may be singular. 
We can define the inverse of a positive semi-definite matrix on $\bar{\R}^{n \times n}$, where $\bar{\R} \triangleq \R \cup \{-\infty, +\infty\}$ is the extended real number line, as follows: 
for $M \succeq 0$, we can write $M = V \Lambda V\t$, where $V$ is orthogonal and $\Lambda = \diag \left[\lambda_{1}, \dots, \lambda_{i}, 0, \dots, 0\right]$ with $\lambda_{1}, \dots, \lambda_{i} > 0$, then, 
its inverse is $M^{-1} \triangleq V \Lambda^{-1} V\t$, where $\Lambda^{-1} \triangleq \diag \left[1/\lambda_{1}, \dots, 1/\lambda_{i}, +\infty, \dots, +\infty\right]$.} 
Consequently, 
\begin{multline*}
\Phi_{A}(0, t) \big(I_{n} - \hat{G}(T, t) \Pi_{t}\big) \hat{G}(T, t) \Phi_{A}(0, t)\t \to 
\\*
W - W (W + \Gamma \Gamma\t)^{-1} W. 
\end{multline*}
From the fact that $\ker (W + \Gamma \Gamma\t) \subseteq \ker W = (\range W)^{\perp}$, it follows that $W (W + \Gamma \Gamma\t)^{-1} W = W (W + \Gamma \Gamma\t)^{+} W$ is well defined, where $M^{+}$ denotes the pseudo-inverse of the matrix $M$. 
Thus, it suffices to show that $z$ is in the left null space of $W - W (W + \Gamma \Gamma\t)^{+} W$, that is, $z\t \big(W - W (W + \Gamma \Gamma\t)^{+} W\big) = 0$.

Let $\rank W = r$ and $\rank [W ~~ \Gamma] = m$, where $r \leq m \leq n$. 
With an appropriate coordinate transformation, we can partition $W$ and $\Gamma$ as follows: 
\begin{align*}
W_{n \times n} &= 
\begin{bmatrix}
W_{1} & 0_{r \times (m-r)} & 0_{r \times (n-m)} \\
0_{(m-r) \times r} & 0_{(m-r) \times (m-r)} & 0_{(m-r) \times (n-m)} \\
0_{(n-m) \times r} & 0_{(n-m) \times (m-r)} & 0_{(n-m) \times (n-m)}
\end{bmatrix}
, \\
\Gamma_{n \times \ell} &= 
\begin{bmatrix}
\Gamma_{1} \\
\Gamma_{2} \\
0_{(n-m) \times \ell}
\end{bmatrix}, 
\end{align*}
where $W_{1} \in \R^{r \times r}$, $W_{1} \succ 0$, $\Gamma_{1} \in \R^{r \times \ell}$, and $\Gamma_{2} \in \R^{(m-r) \times \ell}$. 
In view of Lemma~\ref{lem:invs}, we have that 
\begin{align*}
&W - W (W + \Gamma \Gamma\t)^{+} W 
\\
&= 
\begin{bmatrix}
\Gamma_{1} P_{2} (I_{\ell} + P_{2} \Gamma_{1}\t W_{1}^{-1} \Gamma_{1} P_{2})^{-1} P_{2} \Gamma_{1}\t & 0_{r \times (n-r)} \\
0_{(n-r) \times r} & 0_{(n-r) \times (n-r)}
\end{bmatrix},
\end{align*}
where $P_{2} \triangleq I_{\ell} - \Gamma_{2}\t (\Gamma_{2} \Gamma_{2}\t)^{-1} \Gamma_{2}$ is the orthogonal projection onto $(\range \Gamma_{2}\t)^{\perp}$. 
Let $z \in \R^{n}$ be partitioned in the same way as $\Gamma$. 
That is, let $z\t = [z_{1}\t ~~ z_{2}\t ~~ z_{3}\t]$, where $z_{1} \in \R^{r}$, $z_{2} \in \R^{m-r}$, and $z_{3} \in \R^{n-m}$. 
Since $z\t \Gamma = 0$, we have $z_{1}\t \Gamma_{1} + z_{2}\t \Gamma_{2} = 0$. 
Since $\Gamma_{2} P_{2} = 0$, it follows that $z_{1}\t \Gamma_{1} P_{2} = z_{2}\t \Gamma_{2} P_{2} = 0$. 
We can verify that $z\t \big(W - W (W + \Gamma \Gamma\t)^{+} W\big) = [z_{1}\t \Gamma_{1} P_{2} (I_{\ell} + P_{2} \Gamma_{1}\t W_{1}^{-1} \Gamma_{1} P_{2})^{-1} P_{2} \Gamma_{1}\t ~~ 0_{1 \times (n-r)}] = 0$. 
Since $t \in [0, T]$ is arbitrary, in light of \eqref{eqn:phi-f-phi}, we have shown that, as $\Pi_{0} \to \bar{P}$, then, $\Phi_{A}(0, T) f(\Pi_{0}) \Phi_{A}(0, T)\t$ approaches a matrix for which $z$ lies in its left null space. 
Since $\Phi_{A}(0, T)$ is invertible, it follows that, as the smallest eigenvalue of $\hat{G}(T, 0)^{-1} - \Pi_{0}$ goes to zero, then, $f(\Pi_{0})$ approaches a singular matrix, that is, the smallest eigenvalue of $f(\Pi_{0})$ goes to zero. 
Therefore, for any given $\bar{\Sigma}_{0} \succ 0$, the map $f$ given by \eqref{map:f} is a homomorphism. 
It follows immediately that, for given $\bar{\Sigma}_{0}, \bar{\Sigma}_{T} \succ 0$, the coupled matrix differential equations \eqref{ode:coupled} admit a unique solution $\Pi(t)$. 
Consequently, if $iii)$ holds, then, the optimal control that steers the state covariance of the system~\eqref{sys:cont-time} from $\Sigma(0) = \bar{\Sigma}_{0}$ to $\Sigma(T) = \bar{\Sigma}_{T}$ with the minimum expected control energy exists, and is $u(t) = - B(t)\t \Pi(t) x(t)$. 
Thus, the state covariance of \eqref{sys:cont-time} is controllable on $[0, T]$. 
Therefore, $iii) \implies i)$, which completes the proof. 
\hfill $\qed$


It is well known that the continuous-time matrix pair $\big(A(s), B(s)\big)$ is completely controllable~\cite{silverman1967controllability} on the time interval $[t, T]$ if and only if $G(T, t) \succ 0$~\cite{hendricks2008linear}. 
Similarly to the discrete-time case, we have two corollaries.

\begin{corollary} \label{cor:contr-no-noise-cont}
Assume that there is no additive noise in the system~\eqref{sys:cont-time}, that is, $D(t) \equiv 0$. 
Then, the state covariance of \eqref{sys:cont-time} is controllable on the interval $[0, T]$ if and only if the matrix pair $\big(A(t), B(t)\big)$ is completely controllable on $[0, T]$. 
\end{corollary}

\begin{corollary} \label{cor:contr-all-noise-cont}
Assume that the additive noise in the system~\eqref{sys:cont-time} is non-degenerate, that is, for all $t \in [0, T]$, $\rank D(t) = n$. 
Then, the state covariance of \eqref{sys:cont-time} is controllable on the interval $[0, T]$ if and only if, for all $t \in [0, T)$, the matrix pair $\big(A(s), B(s)\big)$ is completely controllable on $[t, T]$. 
\end{corollary}


\section{Numerical Example} \label{sec:example}

In this section, we first summarize the semi-definite programming (SDP) approaches given in \cite{liu2024mult, liu2024disc} for finding a control that steers the state covariance of either system~\eqref{sys:cont-time} or system~\eqref{sys:disc-time} from a given initial state covariance to a desired reachable state covariance. 
Then, a numerical example is given to illustrate our theoretical results. 
%
%
The SDP method for the continuous-time system~\eqref{sys:cont-time} is described below.

Let $\bar{\Sigma}_{0}$ be the initial state covariance and let $\bar{\Sigma}_{T}$ be the desired terminal state covariance. 
For $t \in [0, T]$, let $Y(t) = \Sigma^{uu}(t)$, let $U(t) = \Sigma^{ux}(t)$, and choose $R(t) \in \R^{p \times p}$, $R(t) \succ 0$. 
The SDP problem is~\cite{liu2024mult}
\begin{align*}
&\min_{\Sigma, U, Y} ~~ \int_{0}^{T} \trace\big(R(t) Y(t)\big) \, \dd t
\\
&
\text{such that}
\\
&
\dot{\Sigma} = A(t) \Sigma + \Sigma A(t)\t  + B(t) U(t)  + U(t)\t B(t)\t 
\\
&\qquad\qquad\qquad\qquad+ D(t) D(t)\t,
\\
&
Y(t) \succeq U(t) \Sigma^{-1}(t) U(t)\t, \quad \Sigma(0) = \bar{\Sigma}_{0}, \quad \Sigma(T) = \bar{\Sigma}_{T}.
\end{align*}
Since $R \succ 0$, it follows that, if $Y_{1} \succeq Y_{2} \succeq 0$, then, 
\begin{align*}
\trace(R Y_{1}) 
&= \trace\big(R^{\frac{1}{2}} Y_{1} R^{\frac{1}{2}}\big)
\\
&\geq \trace\big(R^{\frac{1}{2}} Y_{2} R^{\frac{1}{2}}\big) = \trace(R Y_{2}). 
\end{align*}
In view of the fact that $Y$ does not appear in the equality constraints, we know that the optimal solution of the SDP problem, which minimizes the previous cost function, satisfies $Y(t) = U(t) \Sigma^{-1}(t) U(t)\t$. 
After the above SDP problem is solved, we find $F(t) = U(t) \Sigma^{-1}(t)$. 
In light of Proposition~\ref{prp:stt-fb}, the state feedback control $u(t) = F(t) x(t)$ will steer the state covariance of the system~\eqref{sys:cont-time} from $\bar{\Sigma}_{0}$ to $\bar{\Sigma}_{T}$.


The SDP approach for the discrete-time system~\eqref{sys:disc-time} is summarized as follows. 
More details can be found in~\cite{liu2024disc}. 
Let $\bar{\Sigma}_{0}$ be the given initial state covariance and let $\bar{\Sigma}_{k}$ be the desired terminal state covariance. 
For each $i = 0, 1, \dots, k-1$, let $Y_{i} = \Sigma_{i}^{uu}$, let $U_{i} = \Sigma_{i}^{ux}$, and choose $R_{i} \in \R^{p \times p}$, $R_{i} \succ 0$. 
The SDP problem is~\cite{liu2024disc}
\begin{align*}
&\min_{\Sigma_{i}, U_{i}, Y_{i}} ~~ \sum_{i = 0}^{k-1} \trace \big(R_{i} Y_{i}\big), 
\\
&
\text{such that, for all}\; k = 0, 1, \dots, N-1, 
\\
&
\Sigma_{i+1} = 
A_{i} \Sigma_{i} A_{i}\t + B_{i} U_{i} A_{i}\t + A_{i} U_{i}\t B_{i}\t 
\\
&\qquad\qquad\qquad\qquad\qquad+ B_{i} Y_{i} B_{i}\t + D_{i} D_{i}\t, 
\\
&
\begin{bmatrix} 
\Sigma_{i} & U_{i}\t \\
U_{i} & Y_{i}
\end{bmatrix} 
\succeq 0, \quad 
\Sigma_{0} = \bar{\Sigma}_{0}, \quad \Sigma_{k} = \bar{\Sigma}_{k}. 
\end{align*}
After the above SDP problem is solved, for $i = 0, 1, \dots, k-1$, we find $F_{i} \in R^{p \times n}$ and $V_{i} \in \R^{p \times p}$ such that $U_{i} = F_{i} \Sigma_{i}$ and $V_{i} = Y_{i} - F_{i} \Sigma_{i} F_{i}\t \succeq 0$. 
In light of Proposition~\ref{prp:rand-stt-fb}, the randomized state feedback control $u_{i} = F_{i} x_{i} + \nu_{i}$ will steer the state covariance of the system~\eqref{sys:disc-time} from $\bar{\Sigma}_{0}$ to $\bar{\Sigma}_{k}$, where $\nu_{i} \in \R^{p}$, $i = 0, 1, \dots, k-1$, is a sequence of independent, square-integrable random vectors with $\mathbb{E}\left[\nu_{i}\right] = 0$ and $\mathbb{E}\left[\nu_{i} \nu_{i}\t\right] = V_{i}$. 

Next, we use an example to illustrate our theoretical results. 
Consider the two-dimensional linear stochastic system
\begin{equation} \label{exp:disc-time}
x_{k+1} = A x_{k} + B u_{k} + D w_{k}, \quad k = 0, 1, 2, \dots, 
\end{equation}
where, 
\begin{equation*}
A = 
\begin{bmatrix}
1 & 0.2 \\ 
0 & 0.96
\end{bmatrix}
, \quad
B =
\begin{bmatrix}
1 \\
0
\end{bmatrix}
, \quad
D =
\begin{bmatrix}
0.4 & 0.2 \\
0 & 0.3
\end{bmatrix}. 
\end{equation*}
Note that $(A, B)$ is not controllable and hence the state covariance of the system~\eqref{exp:disc-time} is not controllable from Theorem~\ref{thm:contr-noise}. 
Let $k = 30$. 
According to Theorem~\ref{thm:reach-noise} or Theorem~\ref{thm:reach-noise-2}, all 
reachable state covariance matrices at time $k = 30$ are semi-definite matrices having the (2,2) entry fixed at $1.3079$.
For this example, we therefore choose $\mathbb{E}[x_{0}] = [0 ~~ 0]\t$, $\mathbb{E}[x_{30}] = [30 ~~ 0]\t$, while the initial and  desired terminal state covariances are given by 
\begin{equation*}
\bar{\Sigma}_{0} = 
\begin{bmatrix}
5 & -1 \\ 
-1 & 3
\end{bmatrix}
, \quad
\bar{\Sigma}_{30} = 
\begin{bmatrix}
0.5 & 0.2 \\ 
0.2 & 1.3079
\end{bmatrix}. 
\end{equation*}

\begin{figure}[!h]
\centering
\includegraphics[width=0.46\textwidth]{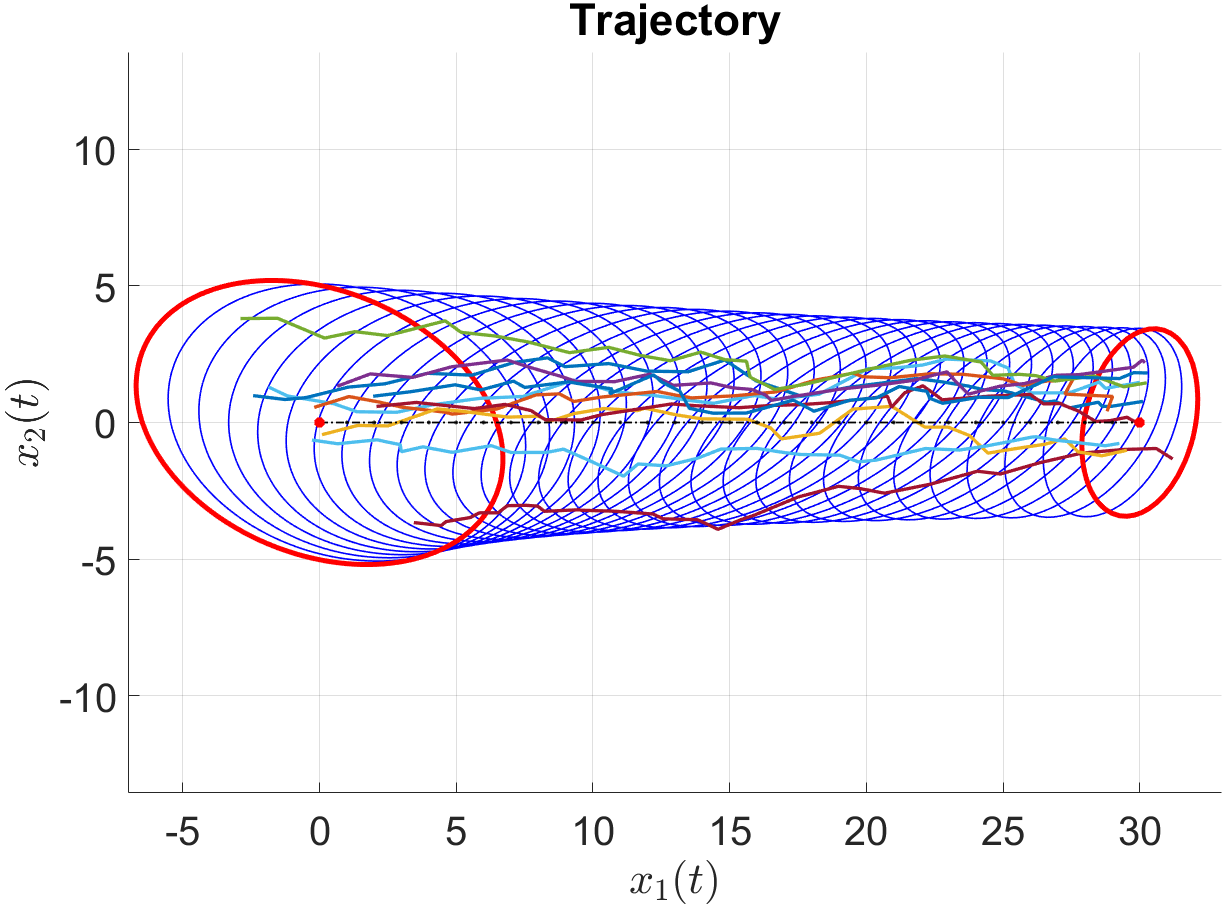}
\caption{Ten sample paths of \eqref{exp:disc-time} and their three-standard-deviation ellipses.}
\label{fig:SDP}
\end{figure}

We choose $R_{i} \equiv 1$ in the cost function of the SDP formulation. 
Figure~\ref{fig:SDP} shows the trajectories of ten sample paths of \eqref{exp:disc-time} and their covariance ellipses, 
where the initial and target three-standard-deviation tolerance regions are marked with red ellipses, 
and the three-standard-deviation tolerance region at each time step is marked with a blue ellipse.
As shown in the figure, the final covariance matrix is achieved even if the set of reachable covariances is not the whole space of positive semi-definite matrices.
%
%


\section{Concluding Remarks}

This paper characterizes the set of reachable terminal state covariances for discrete-time and continuous-time linear stochastic systems with additive noise, starting from a given initial state covariance. 
Specifically, a complete characterization of the reachable set of terminal state covariances 
is given for discrete-time systems, 
while an upper bound on the reachable set of terminal state covariances 
is provided for continuous-time systems. 
We establish necessary and sufficient conditions for the controllability of the state covariance for both discrete-time and continuous-time systems.

A natural extension of this work is to determine the exact set of reachable terminal state covariances for continuous-time systems. 
We also observe a close resemblance between the reachability of the state mean in a stochastic setting (or the reachability of the state in a deterministic setting) and the reachability of the state covariance. 
It would be interesting to find out whether the reachability results for higher moments of the state, if they exist, also bear such a resemblance.

\begin{ack}                               
This work has been supported by NASA University Leadership Initiative award 80NSSC20M0163. 
\end{ack}

\bibliographystyle{myunsrt}
\bibliography{revision}


\appendix

\section{Appendix}

\begin{lemma} \label{lem:sets}
Let $H \in \R^{n \times n}$, $J \in \R^{n \times m}$, and $\bar{\Sigma} \in \R^{n \times n}$, $\bar{\Sigma} \succeq 0$. 
The following sets are equal. 
\begin{enumerate}[leftmargin=*]

\item[i)] $\mathcal{S}_{1} \triangleq \big\{\Sigma \succeq 0 : \Sigma = H \bar{\Sigma} H\t + J X + X\t J\t, X \in \R^{m \times n}\big\}$. 

\item[ii)] $\mathcal{S}_{2} \triangleq \big\{\Sigma \succeq 0 : P \Sigma P = P H \bar{\Sigma} H\t P\big\}$, where $P \in \R^{n \times n}$ is the orthogonal projection onto $(\range J)^{\perp}$.

\item[iii)] $\mathcal{S}_{3} \hspace{-0.5mm} \triangleq \hspace{-0.5mm} \bigg\{\Sigma \succeq 0 \hspace{-0.5mm} : \hspace{-0.5mm}
\Sigma \hspace{-0.5mm} = \hspace{-0.5mm}
\big[H ~~ J\big]
\hspace{-1mm} 
\begin{bmatrix}
\bar{\Sigma} & \bar{\Sigma} F\t \\
F \bar{\Sigma} & F \bar{\Sigma} F\t + V
\end{bmatrix}
\hspace{-1mm} 
\begin{bmatrix}
H\t \\
J\t
\end{bmatrix} \hspace{-1mm} , 
\\
F \in \R^{m \times n}, 
V \in \R^{m \times m}, V \succeq 0
\bigg\}$. 

\item[iv)] $\mathcal{S}_{4} \triangleq \big\{\Sigma \succeq 0 : \Sigma = (H + J F) \bar{\Sigma} (H + J F)\t + J V J\t, F \in \R^{m \times n}, V \in \R^{m \times m}, V \succeq 0\big\}$. 

\end{enumerate}
\end{lemma}

\noindent {\bf Proof:} 
Clearly, $\mathcal{S}_{1} = \mathcal{S}_{2}$, $\mathcal{S}_{3} = \mathcal{S}_{4}$, and $\mathcal{S}_{4} \subseteq \mathcal{S}_{1}$. 
We only need to show that $\mathcal{S}_{1} \subseteq \mathcal{S}_{4}$. 
Let $Q \in \mathcal{S}_{1}$. 
Then, for some $\bar{X} \in \R^{m \times n}$, $Q = H \bar{\Sigma} H\t + J \bar{X} + \bar{X}\t J\t \succeq 0$. 
If $\range H \not\subseteq (\range J)^{\perp}$, we can write $H = H_{1} + H_{2}$, where $\range H_{1} \subseteq (\range J)^{\perp}$ and $\range H_{2} \subseteq \range J$. 
Since $\range H_{2} \subseteq \range J$, there exists $\bar{Y} \in \R^{m \times n}$ such that $H_{2} = J \bar{Y}$. 
It follows that 
\begin{align*}
Q &= (H_{1} + H_{2}) \bar{\Sigma} (H_{1}\t + H_{2}\t) + J \bar{X} + \bar{X}\t J\t
\\
&= H_{1} \bar{\Sigma} H_{1}\t + J \bar{Y} \bar{\Sigma} H_{1}\t + H_{1} \bar{\Sigma} \bar{Y}\t J\t + J \bar{Y} \bar{\Sigma} \bar{Y}\t J\t 
\\
&\hspace{5.1cm}
+ J \bar{X} + \bar{X}\t J\t
\\
&= H_{1} \bar{\Sigma} H_{1}\t + J \bar{Z} + \bar{Z}\t J\t \succeq 0, 
\end{align*}
where $\bar{Z} \triangleq \bar{Y} \bar{\Sigma} H_{1}\t + \bar{X} + \frac{1}{2} \bar{Y} \bar{\Sigma} \bar{Y}\t J\t \in \R^{m \times n}$. 
Without loss of generality, assume that $\range H \subseteq (\range J)^{\perp}$. 
Since $Q \succeq 0$, it follows from Proposition~\ref{prp:pos-semi-defn} that $\range \bar{X}\t \subseteq \range H \bar{\Sigma} H\t + \range J$. 
Hence, there exist matrices $L \in \R^{m \times n}$ and $\bar{W} = \bar{W}\t \in \R^{m \times m}$ such that $Q = H \bar{\Sigma} H\t + J \bar{L} H \bar{\Sigma} H\t + H \bar{\Sigma} H\t \bar{L}\t J\t + J \bar{W} J\t$. 
Since $\range H \subseteq (\range J)^{\perp}$ and $Q \succeq 0$, it follows from Proposition~\ref{prp:pos-semi-defn} that 
\begin{align*}
J \bar{W} J\t 
&\succeq J \bar{L} H \bar{\Sigma} H\t (H \bar{\Sigma} H\t)^{+} H \bar{\Sigma} H\t \bar{L}\t J\t 
\\
&= J \bar{L} H \bar{\Sigma} H\t \bar{L}\t J\t. 
\end{align*}
Thus, there exists a matrix $\bar{V} \succeq 0$ such that $J \bar{W} J\t = J \bar{L} H \bar{\Sigma} H\t \bar{L}\t J\t + J \bar{V} J\t$. 
Let $\bar{F} = \bar{L} H$. 
We then have 
\begin{align*}
Q &= H \bar{\Sigma} H\t \! + \hspace{-0.4mm} J \bar{F} \bar{\Sigma} H\t \! + \hspace{-0.5mm} H \bar{\Sigma} \bar{F}\t J\t \! + \hspace{-0.4mm} J \bar{F} \bar{\Sigma} \bar{F}\t J\t \! + \hspace{-0.4mm} J \bar{V} J\t 
\\
&= (H + J \bar{F}) \bar{\Sigma} (H + J \bar{F})\t + J \bar{V} J\t \in \mathcal{S}_{4}. 
\end{align*}
Therefore, $\mathcal{S}_{1} \subseteq \mathcal{S}_{4}$, which completes the proof. 
\hfill $\qed$


\begin{lemma} \label{lem:sepr}
Let $H \in \R^{n \times n}$, $J \in \R^{n \times m}$, and $\bar{\Sigma}_{1}, \bar{\Sigma}_{2} \in \R^{n \times n}$, $\bar{\Sigma}_{1}, \bar{\Sigma}_{2} \succeq 0$. 
The following two sets are equal. 
\begin{enumerate}[leftmargin=*]

\item[i)] $\mathcal{S}_{5} \triangleq \big\{\Sigma \succeq 0 : \Sigma = (H + J F) (\bar{\Sigma}_{1} + \bar{\Sigma}_{2}) (H + J F)\t + J V J\t, F \in \R^{m \times n}, V \in \R^{m \times m}, V \succeq 0\big\}$. 

\item[ii)] $\mathcal{S}_{6} \triangleq \big\{\Sigma \succeq 0 : \Sigma = (H + J F_{1}) \bar{\Sigma}_{1} (H + J F_{1})\t + J V_{1} J\t + (H + J F_{2}) \bar{\Sigma}_{2} (H + J F_{2})\t + J V_{2} J\t, F_{1}, F_{2} \in \R^{m \times n}, V_{1}, V_{2} \in \R^{m \times m}, V_{1}, V_{2} \succeq 0\big\}$. 

\end{enumerate}
\end{lemma}

\noindent {\bf Proof:} 
Let $Q \in \mathcal{S}_{5}$. 
Then, $Q = (H + J \bar{F}) (\bar{\Sigma}_{1} + \bar{\Sigma}_{2}) (H + J \bar{F})\t + J \bar{V} J\t$ for some $\bar{F} \in \R^{m \times n}$, $\bar{V} \in \R^{m \times m}$, $\bar{V} \succeq 0$. 
Let $\bar{F}_{1} = \bar{F}_{2} = \bar{F}$. 
Let $\bar{V}_{1}, \bar{V}_{2} \in \R^{m \times m}$ be positive semi-definite matrices such that $\bar{V}_{1} + \bar{V}_{2} = \bar{V}$. 
Hence, $Q = (H + J \bar{F}_{1}) \bar{\Sigma}_{1} (H + J \bar{F}_{1})\t + (H + J \bar{F}_{2}) \bar{\Sigma}_{2} (H + J \bar{F}_{2})\t + J \bar{V}_{1} J\t + J \bar{V}_{2} J\t \in \mathcal{S}_{6}$. 
Thus, $\mathcal{S}_{5} \subseteq \mathcal{S}_{6}$. 
Next, we show that $\mathcal{S}_{6} \subseteq \mathcal{S}_{5}$. 
Let $Q \in \mathcal{S}_{6}$. 
Then, for some $\bar{F}_{1}, \bar{F}_{2} \in \R^{m \times n}$, $\bar{V}_{1}, \bar{V}_{2} \in \R^{m \times m}$, $\bar{V}_{1}, \bar{V}_{2} \succeq 0$, 
\begin{multline} \label{eqn:Q}
Q = 
\big[H ~~ J\big]
\Bigg(
\begin{bmatrix}
\bar{\Sigma}_{1} & \bar{\Sigma}_{1} \bar{F}_{1}\t \\
\bar{F}_{1} \bar{\Sigma}_{1} & \bar{F}_{1} \bar{\Sigma}_{1} \bar{F}_{1}\t + \bar{V}_{1}
\end{bmatrix}
\\
+
\begin{bmatrix}
\bar{\Sigma}_{2} & \bar{\Sigma}_{2} \bar{F}_{2}\t \\
\bar{F}_{2} \bar{\Sigma}_{2} & \bar{F}_{2} \bar{\Sigma}_{2} \bar{F}_{2}\t + \bar{V}_{2}
\end{bmatrix}
\Bigg)
\begin{bmatrix}
H\t \\
J\t
\end{bmatrix}.
\end{multline}
Since $\bar{\Sigma}_{1}, \bar{\Sigma}_{2} \succeq 0$, it follows that $\range (\bar{\Sigma}_{1} \bar{F}_{1}\t + \bar{\Sigma}_{2} \bar{F}_{2}\t) \subset \range \bar{\Sigma}_{1} + \range \bar{\Sigma}_{2} = \range (\bar{\Sigma}_{1} + \bar{\Sigma}_{2})$. 
Thus, there exists $\bar{F} \in \R^{m \times n}$ such that $\bar{\Sigma}_{1} \bar{F}_{1}\t + \bar{\Sigma}_{2} \bar{F}_{2}\t = (\bar{\Sigma}_{1} + \bar{\Sigma}_{2}) \bar{F}\t$. 
Since the block matrix in the parenthesis of the right-hand side of \eqref{eqn:Q} is positive semi-definite, it follows from Proposition~\ref{prp:pos-semi-defn} that $\bar{F}_{1} \bar{\Sigma}_{1} \bar{F}_{1}\t + \bar{V}_{1} + \bar{F}_{2} \bar{\Sigma}_{2} \bar{F}_{2}\t + \bar{V}_{2} \succeq \bar{F} (\bar{\Sigma}_{1} + \bar{\Sigma}_{2}) \bar{F}\t$. 
Hence, for some $\bar{V} \in \R^{m \times m}$, $\bar{V} \succeq 0$, we can write
\begin{equation*}
Q \hspace{-0.5mm} = \hspace{-0.5mm} 
\big[H ~~ J\big]
\hspace{-1.5mm}
\begin{bmatrix}
\bar{\Sigma}_{1} + \bar{\Sigma}_{1} & (\bar{\Sigma}_{1} + \bar{\Sigma}_{2}) \bar{F}\t \\
\bar{F}(\bar{\Sigma}_{1} + \bar{\Sigma}_{2}) & \bar{F} (\bar{\Sigma}_{1} + \bar{\Sigma}_{2}) \bar{F}\t + \bar{V}
\end{bmatrix}
\hspace{-2mm}
\begin{bmatrix}
H\t \\
J\t
\end{bmatrix}
\hspace{-1mm}
\in \mathcal{S}_{5}.
\end{equation*}
Thus, $\mathcal{S}_{6} \subseteq \mathcal{S}_{5}$. 
\hfill $\qed$


\begin{lemma} \label{lem:bd}
For $t \in [0, T]$ and $F(t) \in \R^{p \times n}$, let $\Phi_{A+BF}(T, t)$ denote the state transition matrix of $A(\tau) + B(\tau) F(\tau)$ from time $t$ to $T$. 
Then, $\Phi_{A+BF}(T, t) \in \big\{\Phi_{A}(T, t) + G(T, t) K(t) : K(t) \in \R^{n \times n},~\text{with}~ \Phi_{A}(T, t) + G(T, t) K(t) ~\text{invertible}\big\}$, where $G(T, t)$ is given by \eqref{def:G-cont}. 
\end{lemma}

\noindent {\bf Proof:} 
Given $T > 0$, $\Phi_{A+BF}(T, t)$ satisfies 
\begin{align*}
&\dot{\Phi}_{A+BF}(T, t) = - \Phi_{A+BF}(T, t) \big[A(t) + B(t) F(t)\big], 
\\
&\Phi_{A+BF}(T, T) = I_{n}. 
\end{align*}
We can check from direct differentiation that 
\begin{align*}
&\Phi_{A+BF}(T, t) = \Phi_{A}(T, t) 
\\
&\qquad+ \int_{t}^{T} \Phi_{A+BF}(T, \tau) B(\tau) F(\tau) \Phi_{A}(\tau, t) \, \dd \tau.
\end{align*}
We claim that 
\begin{multline} \label{clm:range}
\range \int_{t}^{T} \Phi_{A+BF}(T, \tau) B(\tau) F(\tau) \Phi_{A}(\tau, t) \, \dd \tau 
\\ 
\subset \range G(T, t). 
\end{multline}
Thus, there exists $K(t) \in \R^{n \times n}$, such that 
\begin{equation*}
\Phi_{A+BF}(T, t) = \Phi_{A}(T, t) + G(T, t) K(t), 
\end{equation*}
which is invertible.

To show the claim~\eqref{clm:range} notice that, 
since $G(T, t)$ is symmetric, it follows that $\big(\range G(T, t)\big)^{\perp} = \ker G(T, t)$. 
Hence, to show \eqref{clm:range}, it suffices to show that 
\begin{multline} \label{clm:ker}
\ker G(T, t) \subset 
\\
\left(\range \int_{t}^{T} \Phi_{A+BF}(T, \tau) B(\tau) F(\tau) \Phi_{A}(\tau, t) \, \dd \tau \right)^{\perp}. 
\end{multline}
As the reachability of the pair $\big(A(\tau) + B(\tau) F(\tau), B(\tau)\big)$ is the same as that of $\big(A(\tau), B(\tau)\big)$, 
the reachability Gramian of the pair $\big(A(\tau) + B(\tau) F(\tau), B(\tau)\big)$ has the same range and kernel as that of $\big(A(\tau), B(\tau)\big)$ on $[t, T]$. 
Let $x \in \ker G(T, t)$. 
Then, $x$ also lies in the kernel of the reachability Gramian of 
$\big(A(\tau) + B(\tau) F(\tau), B(\tau)\big)$ on $[t, T]$. 
It follows immediately that 
\begin{equation*}
x\t \int_{t}^{T} \Phi_{A+BF}(T, \tau) B(\tau) B(\tau)\t \Phi_{A+BF}(T, \tau)\t \, \dd \tau \, x = 0.
\end{equation*}
Equivalently, we can write 
\begin{equation*}
\int_{t}^{T} \big\| B(\tau)\t \Phi_{A+BF}(T, \tau)\t x \big\|^{2} \, \dd \tau = 0.
\end{equation*}
It follows that, for $\tau \in [t, T]$, $x\t \Phi_{A+BF}(T, \tau) B(\tau) \equiv 0$. 
Thus, 
\begin{equation*}
x\t \int_{t}^{T} \Phi_{A+BF}(T, \tau) B(\tau) F(\tau) \Phi_{A}(\tau, t) \, \dd \tau = 0. 
\end{equation*}
Hence, 
\begin{equation*}
x \in \left(\range \int_{t}^{T} \Phi_{A+BF}(T, \tau) B(\tau) F(\tau) \Phi_{A}(\tau, t) \, \dd \tau \right)^{\perp}, 
\end{equation*}
and hence~\eqref{clm:ker} holds. 
\hfill $\qed$


\begin{lemma} \label{lem:spec}
Let $M_{1}$ and $M_{2}$ be $n \times n$ real symmetric matrices. 
Then, 
\begin{align*}
\spec \big(M_{1} M_{2}\big) 
= \spec \big(M_{2} M_{1}\big) 
&= \spec \big(M_{1}^{\frac{1}{2}} M_{2} M_{1}^{\frac{1}{2}}\big) 
\\
&= \spec \big(M_{2}^{\frac{1}{2}} M_{1} M_{2}^{\frac{1}{2}}\big), 
\end{align*}
where $\spec(M)$ denotes the spectrum of the matrix $M$, $M^{\frac{1}{2}}$ denotes the square root of $M$. 
Additionally, if $M_{1} \succeq 0$ or $M_{2} \succeq 0$, then $\spec (M_{1} M_{2}) \subset \R$. 
\end{lemma}

\noindent {\bf Proof:} 
Since $M_{1}$ is real symmetric, we can write $M_{1} = P \bar{\Lambda} P\t$, where $P$ is an orthogonal matrix and $\bar{\Lambda}$ is a real diagonal matrix~\cite{horn2012matrix}.
Without loss of generality, assume that $\bar{\Lambda} \triangleq \diag \left[\bar{\lambda}_{1}, \dots, \bar{\lambda}_{k}, 0, \dots, 0\right]$, where $\bar{\lambda}_{1}, \dots, \bar{\lambda}_{k} \neq 0$. 
Let $\varepsilon > 0$, and define $\Gamma_{\varepsilon} \triangleq \diag \left[0, \dots, 0, \varepsilon, \dots, \varepsilon\right]$, where the zero and nonzero diagonal entries of $\Gamma_{\varepsilon}$ complement those of $\bar{\Lambda}$. 
Clearly, $\bar{\Lambda}^{\frac{1}{2}} + \Gamma_{\varepsilon}$ is invertible. 
Let $R_{\varepsilon} \triangleq P \big(\bar{\Lambda}^{\frac{1}{2}} + \Gamma_{\varepsilon}\big) P\t$, which is also invertible. 
Then, for any $\varepsilon > 0$, $M_{1} M_{2}$ is similar to $R_{\varepsilon}^{-1} (M_{1} M_{2}) R_{\varepsilon} = M_{1}^{\frac{1}{2}} M_{2} \big(M_{1}^{\frac{1}{2}} + P \Gamma_{\varepsilon} P\t\big)$. 
Thus, we have $\spec (M_{1} M_{2}) = \spec \big(M_{1}^{\frac{1}{2}} M_{2} (M_{1}^{\frac{1}{2}} + P \Gamma_{\varepsilon} P\t)\big)$. 
It is clear that as $\varepsilon \to 0$, then, $P \Gamma_{\varepsilon} P\t \to 0_{n \times n}$. 
Since the eigenvalues of a matrix are continuous in terms of the values of its entries \cite{harris1987roots}, it follows that, for any $\varepsilon > 0$, $\spec \big(M_{1}^{\frac{1}{2}} M_{2} (M_{1}^{\frac{1}{2}} + P \Gamma_{\varepsilon} P\t)\big) = \spec \big(M_{1}^{\frac{1}{2}} M_{2} M_{1}^{\frac{1}{2}}\big)$. 
Therefore, $\spec (M_{1} M_{2}) = \spec \big(M_{1}^{\frac{1}{2}} M_{2} M_{1}^{\frac{1}{2}}\big)$. 
Since $M_{1}$ and $M_{2}$ are symmetric, we have that $\spec (M_{1} M_{2}) = \spec (M_{2} M_{1})$. 
If $M_{1} \succeq 0$, then $M_{1}^{\frac{1}{2}} M_{2} M_{1}^{\frac{1}{2}}$ is real symmetric, thus has all real eigenvalues. 
\hfill $\qed$










\noindent {\bf Proof of Theorem~\ref{thm:pi-exist}:} 
We know from~\cite{kilicaslan2010existence} that \eqref{ode:pi-1} with initial condition $\Pi(0)$ admits the unique solution \eqref{sol:pi} on $[0, T]$ if and only if, for all $t \in [0, T]$, $I_{n} - \hat{G}(t, 0) \Pi(0)$ is invertible. 
It follows immediately from Lemma~\ref{lem:spec}, that the spectrum of $I_{n} - \hat{G}(t, 0) \Pi(0)$ equals that of $I_{n} - \hat{G}(t, 0)^{\frac{1}{2}} \Pi(0) \hat{G}(t, 0)^{\frac{1}{2}}$. 
Since, at $t = 0$, $I_{n} - \hat{G}(0, 0)^{\frac{1}{2}} \Pi(0) \hat{G}(0, 0)^{\frac{1}{2}} = I_{n} \succ 0$, it suffices to show that, for all $t \in [0, T]$, $I_{n} - \hat{G}(t, 0)^{\frac{1}{2}} \Pi(0) \hat{G}(t, 0)^{\frac{1}{2}} \succ 0$ if and only if $I_{n} - \hat{G}(T, 0)^{\frac{1}{2}} \Pi(0) \hat{G}(T, 0)^{\frac{1}{2}} \succ 0$. 
The necessity is obvious. 
We now show the sufficiency.

Let $\rank \hat{G}(T, 0) = r \leq n$. 
Let $L \triangleq [L_{1} ~~ L_{2}] \in \R^{n \times n}$ be an orthogonal matrix\footnote{Notice that such a matrix $L$ is not unique. Any choice of such $L$ will do.} with $L_{1} \in \R^{n \times r}$, $L_{2} \in \R^{n \times (n-r)}$, $\range L_{1} = \range \hat{G}(T, 0)$, and $\range L_{2} = \ker \hat{G}(T, 0)$. 
Since $\range \hat{G}(T, 0)^{\frac{1}{2}} = \range \hat{G}(T, 0)$ and $L L\t = L_{1} L_{1}\t + L_{2} L_{2}\t = I_{n}$, multiplying $I_{n} - \hat{G}(T, 0)^{\frac{1}{2}} \Pi(0) \hat{G}(T, 0)^{\frac{1}{2}} \succ 0$ by $L\t$ on the left and by $L$ on the right yields 
\begin{equation*}
I_{r} - L_{1}\t \hat{G}(T, 0)^{\frac{1}{2}} L_{1} L_{1}\t \Pi(0) L_{1} L_{1}\t \hat{G}(T, 0)^{\frac{1}{2}} L_{1} \succ 0. 
\end{equation*}
Since $L_{1}\t \hat{G}(T, 0)^{\frac{1}{2}} L_{1} \succ 0$, it follows immediately from the above inequality that 
\begin{equation} \label{ineq:pi0-GT}
L_{1}\t \Pi(0) L_{1} \prec \big(L_{1}\t \hat{G}(T, 0)^{\frac{1}{2}} L_{1}\big)^{-2}. 
\end{equation}
Since, for all $t \in [0, T]$, $\hat{G}(t, 0) \preceq \hat{G}(T, 0)$, it is clear that $L_{1}\t \hat{G}(t, 0) L_{1} \preceq L_{1}\t \hat{G}(T, 0) L_{1}$. 
Since $\range \hat{G}(t, 0)^{\frac{1}{2}} = \range \hat{G}(t, 0) \subseteq \range \hat{G}(T, 0)$, we have $\hat{G}(t, 0)^{\frac{1}{2}} L_{2} = 0$ and 
\begin{align*}
\hat{G}(t, 0) = \hat{G}(t, 0)^{\frac{1}{2}} \hat{G}(t, 0)^{\frac{1}{2}} 
&= \hat{G}(t, 0)^{\frac{1}{2}} L L\t \hat{G}(t, 0)^{\frac{1}{2}} 
\\
&= \hat{G}(t, 0)^{\frac{1}{2}} L_{1} L_{1}\t \hat{G}(t, 0)^{\frac{1}{2}}. 
\end{align*}
Then, it follows from $L_{1}\t \hat{G}(t, 0) L_{1} \preceq L_{1}\t \hat{G}(T, 0) L_{1}$ and the above equation that 
$\big(L_{1}\t \hat{G}(t, 0)^{\frac{1}{2}} L_{1}\big)^{2} \preceq \big(L_{1}\t \hat{G}(T, 0)^{\frac{1}{2}} L_{1}\big)^{2}$. 
Since $L_{1}\t \hat{G}(T, 0)^{\frac{1}{2}} L_{1} \succ 0$, we know that the largest singular value of 
\begin{equation*}
\big(L_{1}\t \hat{G}(T, 0)^{\frac{1}{2}} L_{1}\big)^{-1} \big(L_{1}\t \hat{G}(t, 0)^{\frac{1}{2}} L_{1}\big)
\end{equation*} 
is no greater than $1$~\cite{horn2012matrix}. 
Thus, we have that 
\begin{equation} \label{ineq:Gt-GT}
\big(L_{1}\t \hat{G}(t, 0)^{\frac{1}{2}} L_{1}\big) \big(L_{1}\t \hat{G}(T, 0)^{\frac{1}{2}} L_{1}\big)^{-2} \big(L_{1}\t \hat{G}(t, 0)^{\frac{1}{2}} L_{1}\big) \! \preceq I_{r}. 
\end{equation}
To show that, $I_{n} - \hat{G}(t, 0)^{\frac{1}{2}} \Pi(0) \hat{G}(t, 0)^{\frac{1}{2}} \succ 0$, for all $t \in [0, T]$,  
it suffices to show that $L\t \big(I_{n} - \hat{G}(t, 0)^{\frac{1}{2}} \Pi(0) \hat{G}(t, 0)^{\frac{1}{2}}\big) L \succ 0$. 
Since $\hat{G}(t, 0)^{\frac{1}{2}} L_{2} = 0$ and $\hat{G}(t, 0)^{\frac{1}{2}} = \hat{G}(t, 0)^{\frac{1}{2}} L L\t = \hat{G}(t, 0)^{\frac{1}{2}} L_{1} L_{1}\t$, it suffices to show that, for all $t \in [0, T]$, $I_{r} - L_{1}\t \hat{G}(t, 0)^{\frac{1}{2}} L_{1} L_{1}\t \Pi(0) L_{1} L_{1}\t \hat{G}(t, 0)^{\frac{1}{2}} L_{1} \succ 0$.

Let $z \in \R^{r}$ be a nonzero vector. 
Then, we can write $z = z_{1} + z_{2}$, where $z_{1} \in \range \big(L_{1}\t \hat{G}(t, 0)^{\frac{1}{2}} L_{1}\big)$, $z_{2} \in \ker \big(L_{1}\t \hat{G}(t, 0)^{\frac{1}{2}} L_{1}\big)$, $z_{1}\t z_{2} = 0$, $z_{1} \neq 0$ or $z_{2} \neq 0$. 
If $z_{1} = 0$, 
$z\t \big(I_{r} - L_{1}\t \hat{G}(t, 0)^{\frac{1}{2}} L_{1} L_{1}\t \Pi(0) L_{1} L_{1}\t \hat{G}(t, 0)^{\frac{1}{2}} L_{1}\big) z 
= z_{2}\t z_{2} > 0$. 
If $z_{1} \neq 0$, it follows from \eqref{ineq:pi0-GT}, \eqref{ineq:Gt-GT}, and $L_{1}\t \hat{G}(t, 0)^{\frac{1}{2}} L_{1} z_{1} \neq 0$ that 
\begin{align*}
&z\t \big(I_{r} - L_{1}\t \hat{G}(t, 0)^{\frac{1}{2}} L_{1} L_{1}\t \Pi(0) L_{1} L_{1}\t \hat{G}(t, 0)^{\frac{1}{2}} L_{1}\big) z 
\\
&= z_{1}\t z_{1} + z_{2}\t z_{2} 
\\
&\quad- z_{1}\t L_{1}\t \hat{G}(t, 0)^{\frac{1}{2}} L_{1} L_{1}\t \Pi(0) L_{1} L_{1}\t \hat{G}(t, 0)^{\frac{1}{2}} L_{1} z_{1}
\\
&> z_{1}\t z_{1} + z_{2}\t z_{2} 
\\*
&\hspace{3mm}
- z_{1}\t L_{1}\t \hat{G}(t, 0)^{\frac{1}{2}} L_{1} \big(L_{1}\t \hat{G}(T, 0)^{\frac{1}{2}} L_{1}\big)^{-2} L_{1}\t \hat{G}(t, 0)^{\frac{1}{2}} L_{1} z_{1}
\\
&\geq z_{1}\t z_{1} + z_{2}\t z_{2} - z_{1}\t z_{1} = z_{2}\t z_{2} \geq 0.
\end{align*}
Therefore, $I_{r} - L_{1}\t \hat{G}(t, 0)^{\frac{1}{2}} L_{1} L_{1}\t \Pi(0) L_{1} L_{1}\t \hat{G}(t, 0)^{\frac{1}{2}} L_{1} \succ 0$, for all $t \in [0, T]$,
thus completing the proof. 
\hfill $\qed$


\begin{lemma} \label{lem:invs}
Let $W \in \R^{m \times m}$, $W \succeq 0$ having rank $r$, and let $\Gamma \in \R^{m \times \ell}$ be such that $\rank [W ~~ \Gamma] = m$, where $r \leq m \leq r + \ell$. 
With a suitable coordinate transformation, we can write 
$
W = 
\begin{bmatrix}
W_{1} & 0 \\
0 & 0
\end{bmatrix}
$ 
and 
$\Gamma = 
\begin{bmatrix}
\Gamma_{1} \\
\Gamma_{2}
\end{bmatrix}
$, 
where $W_{1} \in \R^{r \times r}$, $W_{1} \succ 0$, $\Gamma_{1} \in \R^{r \times \ell}$, $\Gamma_{2} \in \R^{(m-r) \times \ell}$, and $\rank \Gamma_{2} = m - r$. 
Let $P_{2} \triangleq I_{\ell} - \Gamma_{2}\t (\Gamma_{2} \Gamma_{2}\t)^{-1} \Gamma_{2}$ be the orthogonal projection along $\range \Gamma_{2}\t$ onto its orthogonal complement in $\R^{\ell}$. 
Then, we have 
\begin{multline*}
W (W + \Gamma \Gamma\t)^{-1} W = 
\\
\begin{bmatrix}
W_{1} - \Gamma_{1} P_{2} (I_{\ell} + P_{2} \Gamma_{1}\t W_{1}^{-1} \Gamma_{1} P_{2})^{-1} P_{2} \Gamma_{1}\t & 0 \\
0 & 0
\end{bmatrix}.
\end{multline*}
\end{lemma}

\noindent {\bf Proof:} 
Since $\rank [W ~~ \Gamma] = m$, we have $W + \Gamma \Gamma\t \succ 0$ and $\Gamma_{2}$ has full row rank. 
Since 
$
W + \Gamma \Gamma\t = 
\begin{bmatrix}
W_{1} + \Gamma_{1} \Gamma_{1}\t & \Gamma_{1} \Gamma_{2}\t \\
\Gamma_{2} \Gamma_{1}\t & \Gamma_{2} \Gamma_{2}\t
\end{bmatrix}
$, 
it follows that~\cite{horn2012matrix} 
\begin{multline*}
(W + \Gamma \Gamma\t)^{-1} = 
\\
\begin{bmatrix}
\big(W_{1} + \Gamma_{1} \Gamma_{1}\t - \Gamma_{1} \Gamma_{2}\t (\Gamma_{2} \Gamma_{2}\t)^{-1} \Gamma_{2} \Gamma_{1}\t\big)^{-1} & \star \\
\star & \star 
\end{bmatrix}, 
\end{multline*}
where the $\star$ entries are irrelevant. 
Since $\Gamma_{1} \Gamma_{1}\t - \Gamma_{1} \Gamma_{2}\t (\Gamma_{2} \Gamma_{2}\t)^{-1} \Gamma_{2} \Gamma_{1}\t = \Gamma_{1} \big(I_{\ell} - \Gamma_{2}\t (\Gamma_{2} \Gamma_{2}\t)^{-1} \Gamma_{2}\big) \Gamma_{1}\t = \Gamma_{1} P_{2} \Gamma_{1}\t = \Gamma_{1} P_{2}^{2} \Gamma_{1}\t$, it follows from the Woodbury matrix identity~\cite{horn2012matrix} that $(W_{1} + \Gamma_{1} P_{2}^{2} \Gamma_{1}\t)^{-1} = W_{1}^{-1} - W_{1}^{-1} \Gamma_{1} P_{2} (I_{\ell} + P_{2} \Gamma_{1}\t W_{1}^{-1} \Gamma_{1} P_{2})^{-1} P_{2} \Gamma_{1}\t W_{1}^{-1}$. 
Thus, 
\begin{multline*}
W (W + \Gamma \Gamma\t)^{-1} W = 
\\
\begin{bmatrix}
W_{1} - \Gamma_{1} P_{2} (I_{\ell} + P_{2} \Gamma_{1}\t W_{1}^{-1} \Gamma_{1} P_{2})^{-1} P_{2} \Gamma_{1}\t & 0 \\
0 & 0
\end{bmatrix}.
\end{multline*}
This completes the proof. 
\hfill $\qed$


\end{document}